\documentclass[compsoc, conference, a4paper, 10pt, times]{IEEEtran}

\usepackage{cite}
\usepackage{amsmath,amssymb,amsfonts}
\usepackage{algorithmic}
\usepackage{graphicx}
\usepackage{textcomp}
\usepackage{xcolor}
\def\BibTeX{{\rm B\kern-.05em{\sc i\kern-.025em b}\kern-.08em
    T\kern-.1667em\lower.7ex\hbox{E}\kern-.125emX}}

\usepackage{tikz}

\newlength{\bibitemsep}
\newlength{\bibparskip}
\let\oldthebibliography\thebibliography
\renewcommand\thebibliography[1]{%
  \oldthebibliography{#1}%
  \setlength{\parskip}{\bibitemsep}%
  \setlength{\itemsep}{\bibparskip}%
}

\usepackage[normalem]{ulem}

\newcommand{\myparagraph}[1]{\vspace{1ex}\noindent{\bf #1}}
\makeatletter

\renewcommand*{\@fnsymbol}[1]{\ensuremath{\ifcase#1\or *\or \dagger\or \ddagger\or
   \mathsection\or \mathparagraph\or \|\or **\or \dagger\dagger
   \or \ddagger\ddagger \else\@ctrerr\fi}}
   
\DeclareRobustCommand\onedot{\futurelet\@let@token\@onedot}
\def\@onedot{\ifx\@let@token.\else.\null\fi\xspace}
\def\eg{\emph{e.g}\onedot} 
\def\ie{\emph{i.e}\onedot}

\def\etal{\emph{et al}\onedot}
\usepackage{pifont}
\makeatother
\usepackage{amsmath}
\usepackage{makecell}

\usepackage{hhline}
\usepackage{etoolbox}
\expandafter\patchcmd\csname citeauthor \endcsname
  {\begingroup}{\begingroup\aftergroup\@}{}{}

\usepackage{amsmath,amssymb,amsfonts}
\usepackage{mathtools}
\usepackage{booktabs}
\usepackage[algoruled,nofillcomment,algo2e]{algorithm2e}
\usepackage{amsthm}
\usepackage{caption} %
\usepackage{subcaption} %
\usepackage{bbm} %

\usepackage{hyperref}
\hypersetup{breaklinks=true}

\usepackage{cleveref}
\usepackage{xspace} %
\usepackage{multirow}

\usepackage{cuted}

\usepackage{lastpage} %

\theoremstyle{definition}
\newtheorem{df}{Definition}

\def\namenoformat{{trap weights}\xspace}

\DeclareRobustCommand\encircle[1]{\tikz[baseline=(char.base)]{\node[shape=circle,fill=black,inner sep=1pt] (char) {\textcolor{white}{#1}}}}

\newcommand{\user}{user\xspace}
\newcommand{\users}{users\xspace}
\newcommand{\User}{User\xspace}
\newcommand{\Users}{Users\xspace}
\newcommand{\cp}{server\xspace}

\newcommand{\Cps}{Servers\xspace}

\newcommand{\attackers}{attackers\xspace}

\newcommand{\SA}{SA\xspace}

\newcommand{\WAll}{\mathcal{W}} %
\newcommand{\att}[1]{^{[#1]}} %
\newcommand{\Ltwo}{\ell_2} %
\newcommand{\totalusers}{\mathsf{N}}
\newcommand{\selectusers}{\mathsf{M}}

\newcommand{\rev}[1]{{\color{black}{}#1}}

\begin{document}

\title{Reconstructing Individual Data Points in Federated Learning \\ Hardened with Differential Privacy and Secure Aggregation$^\diamond$}

\author{
\IEEEauthorblockN{%
    Franziska Boenisch\IEEEauthorrefmark{1},
    Adam Dziedzic\IEEEauthorrefmark{1}\IEEEauthorrefmark{2}\textsuperscript{\textsection},
    Roei Schuster\IEEEauthorrefmark{1}\textsuperscript{\textsection},\\
    Ali Shahin Shamsabadi\IEEEauthorrefmark{1}\IEEEauthorrefmark{3}\textsuperscript{\textsection},
    Ilia Shumailov\IEEEauthorrefmark{1}\textsuperscript{\textsection}, and
    Nicolas Papernot\IEEEauthorrefmark{1}\IEEEauthorrefmark{2}
  }%
  \IEEEauthorblockA{\IEEEauthorrefmark{1}Vector Institute \IEEEauthorrefmark{2}University of Toronto \IEEEauthorrefmark{3}The Alan Turing Institute}%
}

\maketitle

\begingroup\renewcommand\thefootnote{\textsection}
\footnotetext{Equal contribution. \newline \indent $^\diamond$. Accepted at the 8th IEEE European Symposium on Security and Privacy (IEEE Euro S\&P).}
\endgroup

\begin{abstract}
Federated learning (FL) is a framework for users to jointly train a machine learning model. 
FL is promoted as a privacy-enhancing technology (PET) that provides data minimization: data never “leaves” personal devices and users share only model updates with a server (e.g., a company) coordinating the distributed training. 
While prior work showed that in vanilla FL a malicious server can extract users' private data from the model updates, in this work we take it further and demonstrate that a malicious server can reconstruct user data even in hardened versions of the protocol.
More precisely, we propose an attack against FL protected with distributed differential privacy (DDP) and secure aggregation (SA). 
Our attack method is based on the introduction of sybil devices that deviate from the protocol to expose individual users' data for reconstruction by the server. 
The underlying root cause for the vulnerability to our attack is a power imbalance: 
the server orchestrates the whole protocol and users are given little guarantees about the selection of other users participating in the protocol. 
Moving forward, we discuss requirements for privacy guarantees in FL.
We conclude that users should only participate in the protocol when they trust the server or they apply local primitives such as local DP, shifting power away from the server.
Yet, the latter approaches come at significant overhead in terms of performance degradation of the trained model, making them less likely to be deployed in practice.

\end{abstract}

\section{Introduction}
\label{sec:introduction}

Federated Learning (FL)~\cite{McMahan.2017Communication} is a widely deployed protocol for collaborative machine learning (ML).
It allows a \cp to train an ML model on data of different \users without requiring direct access to that data.
For this reason, FL is often promoted as data minimizing~\cite{googleBlogpost2022}:
instead of sharing their data, \users calculate model updates, usually gradients, on a shared model obtained from the \cp.
These model updates are then aggregated and applied iteratively to train this shared model.

Prior work has demonstrated that the model updates leak sensitive information on the \users' local training data~\cite{Hitaj.2017Deep,Phong.2017Privacy}. 
This is to be expected since nothing in the design of FL prevents information leakage.
Indeed, the root cause of why FL is inherently vulnerable to data reconstruction attacks is that it is designed to provide confidentiality (data does not leave user devices) rather than privacy (outputs of the computation do not leak sensitive attributes from the \users' input). Without additional privacy measures, FL cannot protect \users from the \cp reconstructing their data. 

Fortunately, \emph{if} the \cp is trusted to follow the protocol as prescribed, a relatively cost-effective mitigation exists: 
the \cp can add noise to the model updates and, thereby, implement differential privacy (DP)~\cite{Dwork.2006Differential} during the aggregation~\cite{mcmahan2018learning}. This degrades the performance of learned models~\cite{wei2020federated}, but adds strong privacy guarantees. %
Unfortunately, the \cp cannot always be trusted. Worse yet, by observing local \users' model updates, an untrusted \cp can mount powerful attacks to reconstruct \users' training data points~\cite{Geiping.2020Inverting,Wang.2019Beyond,Yin.2021See,Zhu.2020Deep,Zhao.2020iDLG, boenisch2021curious,fowl2021robbing}. Contemporary attacks exploit the fact that neurons compute a weighted sum of their input; thus the corresponding gradients contain rescaled versions of the input.

\begin{figure*}[tb]
\centering
\includegraphics[width=0.88\textwidth, trim={2.5cm 6.5cm 2cm 3.5cm},clip]{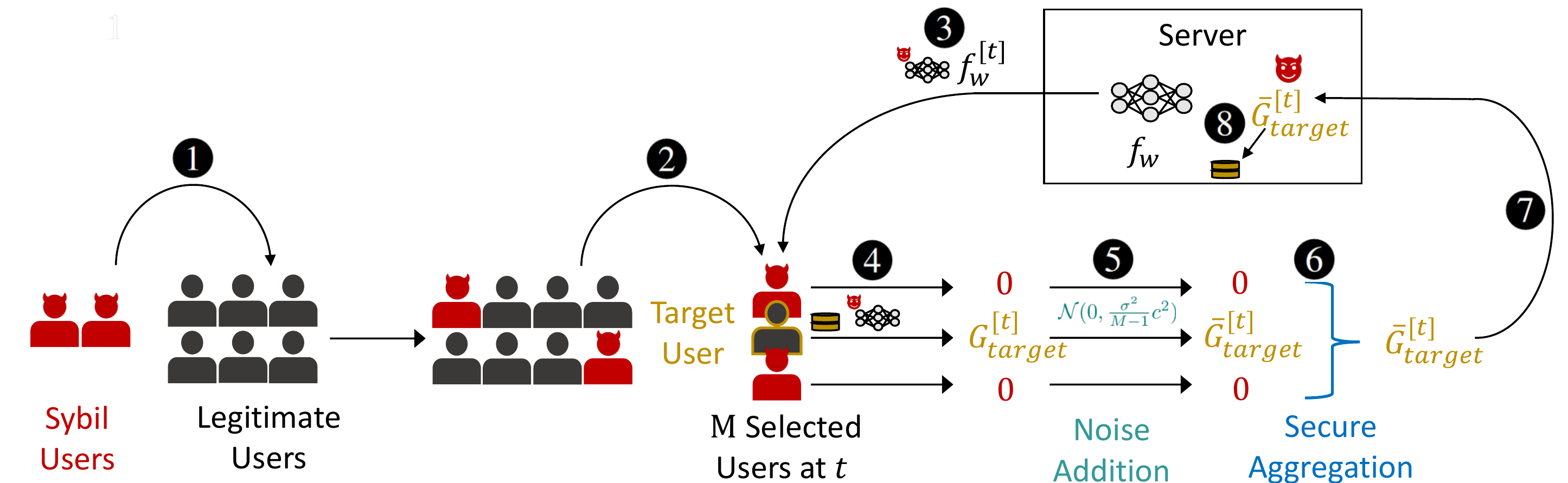}
\vspace{0.4cm}
\caption{\textbf{Course of our attack against FL protected by \SA and DDP.} \encircle{1} The \cp introduces a small fraction of sybil \users into the FL application. \encircle{2} The \cp selects $\selectusers$ \users for participation in training round $t$: one target \user and $\selectusers-1$  sybils. \encircle{3} The \cp manipulates the shared model with trap weights~\cite{boenisch2021curious} and sends it out to the selected \users. \encircle{4} The target \user locally calculates its gradients on the manipulated model while the sybil \users return zero or constant value gradients that are known by the server. %
\encircle{5} Only the target \user locally applies a small amount of noise to its gradients to implement DDP. \encircle{6} The target \user's local noised gradients are aggregated with the sybil \users' values. \encircle{7} The resulting aggregate which effectively contains solely the target \user's gradients is sent to the \cp. 
\encircle{8} The \cp extracts the target \user's training data from the received gradients.
}
\label{fig:FL+SA+DDP}
\end{figure*}

We go a step beyond these prior extraction attacks that operate in vanilla FL protocols and mount an attack against FL combined with secure aggregation (\SA)~\cite{bonawitz2017practical} and distributed differential privacy (DDP)~\cite{truex2019hybrid}.
In \SA, gradient aggregation is performed via a decentralized multiparty computation protocol. In DDP, each \user adds a small amount of noise to their gradient updates. Through the aggregation of \user updates, the cumulative noise of DDP provides sufficiently high privacy guarantees to protect \user data from leaking sensitive information.
The two techniques were designed to decrease the amount of trust placed by users in the central party in FL~\cite{agarwal2021skellam,kairouz2021distributed,girgis2021shuffled}.
Yet, our attack (see \Cref{fig:FL+SA+DDP}) shows that an untrusted \cp that exerts their full capabilities is able to still reconstruct individual \user data points in this setup.

To be clear, we are, of course, not claiming that the cryptographic primitives behind \SA and DDP are broken.
We merely notice that the trust model that they assume, where in each round enough honest \users contribute noised gradient updates, is not necessarily realized in practice within FL. 
In reality, the \cp is entrusted with \emph{provisioning} \users and \emph{sampling} them in each round.
A powerful untrusted \cp can inject an arbitrary number of malicious sybils under their control into any round, as demonstrated by industry actors actively deploying FL~\cite{ramaswamy2020training}.
By sampling a target \user together with a group of sybils that act maliciously, the \cp can acquire direct access to the target \user's non-aggregated update. 
This allows the \cp to eliminate any effect of \SA, and effectively reduces the protection of DDP to a minimum because the noise added by a single \user is not designed to offer the claimed privacy guarantees.

Next, we observe that the \cp is also typically entrusted with \emph{controlling the shared model}.
Prior work~\cite{fowl2021robbing, boenisch2021curious, wen2022fishing} has shown that this ability enables the server to extract large amounts of the \users' individual training data points from gradients in vanilla FL.
By integrating the trap weight approach from~\cite{boenisch2021curious} into our attack, we show that an untrusted \cp can extract high-fidelity \user data points for common learning tasks despite DDP and \SA.
This highlights that \emph{even elaborate combinations of techniques like DDP and \SA with FL can be attacked to perform data reconstruction when the \cp's real-world capabilities are taken into account}.

While exploring data reconstruction under DP, we make another observation that advances our understanding of which users are more vulnerable to data reconstruction attacks: the noise addition does not guarantee equal protection over all model gradients.
As we visualize in \Cref{fig:dp-before-cluster}, some data points can be extracted with higher fidelity than others. This is despite the fact that all corresponding gradients were protected with the same clipping and noising operations.
We identify the gradient norm as the reason behind disparate protection.
For small-norm gradients, noise dominates the signal of the extracted data more than %
for gradients with a large norm.
In principle, clipping in DP for ML is supposed to bound the gradient norm to control for this.
However, the gradient norm is calculated globally %
whereas data is extracted locally from the components of the gradient that correspond to the weighted input of a single neuron. %
When the gradient corresponding to a neuron is large but all other neurons in that layer have small gradients, the overall gradient can still be below the clipping norm.
Therefore, no clipping is performed, the same amount of noise is added to neurons with large and small gradients, and data from the neurons with larger gradients can be extracted with higher fidelity. 
We also sketch a construction that amplifies this effect and makes a few individual data points nearly perfectly extractable, even under noise addition.

We then discuss the centralization and resulting power imbalance between the \cp and \users as the root cause of FL's vulnerability against attacks like the one proposed in this work.
This motivates us to explore the requirements for building a variant of FL that practically prevents attacks by a malicious \cp.
We consider three approaches based on (1) decentralization, (2) user verification, and (3) the support of specialized hardware.
We find that one promising direction is to add adequate amounts of noise to \users' gradients via a cryptographic protocol such as secure multiparty computation~(SMCP).
However as of yet, due to the gradients' high dimensionality, known constructions' communication costs are prohibitive.

As an alternative direction, \users in FL can take responsibility for implementing their full privacy protection locally, for example, by adding enough noise to individually implement strong privacy guarantees.
This approach is commonly referred to as local DP (LDP).
As a last resort, \users can decide not to participate in FL protocols if they do not trust the \cp.
However, the last two options are only available if \users have the required control over their participation in the protocol, which is not always the case.

In summary, we make the following contributions:
\begin{itemize}
    \item We design an attack that enables a malicious \cp~that holds the power of introducing sybil devices to reconstruct individual training data points from \users when FL is protected by \SA and DDP. See \Cref{fig:FL+SA+DDP} for an illustration of our attack flow.
    
    \item We experimentally validate the attack's ability to reconstruct image and textual data with high fidelity in different DDP setups. We propose a proof of concept construction that allows to increase the fidelity of the reconstruction of \users' data points by reducing the effect of additive noise at the level of individual neurons.
    We thus observe disparate leakage over gradients under DDP. 
    
    \item We discuss centralization in FL and its resulting power-imbalance between the \cp and the \users as the root cause of FL's vulnerability and consider potential mitigations.

\end{itemize}

\setcounter{footnote}{0}

\section{Background}
\label{sec:threat_model_and_system}
This section provides background on FL, describes \user-data extraction from gradients in the vanilla version of the protocol and introduces SA and DP---extensions implementing dedicated defenses against privacy leakage.

\subsection{Federated Learning}
\label{sec:background}
FL~\cite{McMahan.2017Communication} is a communication protocol that allows a group of $\totalusers$ \users to jointly train an ML model $f$, such that data never leaves the respective \users' devices. FL involves a \textit{\cp}, who coordinates the training in an iterative process, as follows: at round $t=0$, the \cp initializes the shared model $\WAll\att{0}$ at random, typically following common weight-initialization practice~\cite{Giryes.2016Deep,Glorot.2011Deep,He.2015Delving}. At every round $t$, the \cp chooses a subset of $\selectusers \leq \totalusers$ \users to contribute to the learning round. The \cp then sends each of these \users the model $f_{\WAll\att{t}}$. Each user chooses a subsample of their training data termed a \textit{mini-batch}, computes the gradients of an objective function over $f_{\WAll\att{t}}$ for each of these samples, and returns these gradients, which we call a (local)  \textit{update}, to the \cp. To conclude the round, the \cp updates the shared model by aggregating the received gradients, multiplying them by a learning-rate parameter and applying the change to the shared model.\footnote{For readers unfamiliar with gradient optimization: such gradient-based weight updates are intended to prescribe which direction $\WAll\att{t}$ needs to move towards, for the objective to be minimized. Typically, the objective is a value measuring the model's level of prediction error across a given mini-batch.}

It follows from the description above that \users' training data significantly affects the values of local updates.
This enables a variety of privacy attacks that extract individual \user-data points directly from the model updates, as highlighted in the following section.

\subsection{Data Extraction from Vanilla FL}
\label{sec:attacking_vanilla_FL}

Prior work~\cite{fowl2021robbing, boenisch2021curious, wen2022fishing} has shown that an untrusted \cp can directly extract \user-data from the model gradients. 
In these attacks, the \cp leverages its control over the shared model.\footnote{There also exist optimization-based data reconstruction attacks operating on model updates.
These attacks can be conducted by a passive attacker solely observing the gradients. 
However, computation is expensive and the reconstructed data is not necessarily of high-fidelity.
We provide a brief overview of this type of attack in \Cref{sub:passive_extraction}.}
In \cite{fowl2021robbing}, the \cp exploits this ability to insert a fully-connected model layer as an extraction module where individual data points can be directly extracted.
\cite{boenisch2021curious} manipulates the model weights with an attack they call \namenoformat.
This attack increases natural data leakage from fully-connected model layers.
In \cite{wen2022fishing}, the \cp instead modifies model parameters to extract single data points by increasing their gradient contribution.
The attack operates in several iterations of the protocol to collect multiple gradient updates from the same user.

The presence of such data extraction attacks highlights the vulnerability of vanilla FL to privacy-leakage.
To prevent training data from leaking onto updates and straight to the hands of the \cp, various extensions have been proposed, as we now briefly describe.

\subsection{Secure Aggregation}
\label{sub:sa}
In \SA, due to Bonawitz et al.~\cite{bonawitz2017practical}, \users do not send their individual updates to the \cp. Instead, they perform, along with the \cp, a multiparty computation (MPC) protocol that ensures the \cp only receives the average of all updates in the round.
Various improvements of the original protocol were suggested, for example, to allow the \cp to prove the correctness of the aggregate computation~\cite{xu2019verifynet}, 
increase robustness against malicious \users' manipulated gradients~\cite{burkhalter2021rofl}, or improve communication efficiency~\cite{bell2020secure, guo2020v}.

\subsection{Differential Privacy in FL}
\label{sub:dp}
Nothing in the design of FL prevents information leakage: FL is designed to provide confidentiality (data does not leave user devices) rather than privacy (outputs of the computation do not leak sensitive attributes from the \users' input). 
As discussed in \Cref{sec:attacking_vanilla_FL}, this leaves vanilla FL vulnerable to data reconstruction attacks.
To ensure the privacy of \users' sensitive training data, it is natural to consider DP approaches that work by adding noise to \user updates. DP is a gold standard in privacy technology because proper application of it comes with a theoretical bound on the probability of an adversary being able to distinguish adjacent datasets, \ie datasets that differ solely in one data point.
This implies a bound on the probability of data point extraction. 
In other words, a DP approach properly applied to FL updates could, in principle, ensure that individual \user data points are not revealed to whoever observes the noised updates. See Appendix~\ref{sec:app_DP} for more background on DP and its integration to ML.

One possible approach to integrate DP into FL is \textit{centralized DP} (CDP)~\cite{ramaswamy2020training, balle2020privacy, kairouz2021practical}, where the \cp adds noise to the mini-batch gradients received by \users. CDP assumes that the \cp is trusted to add noise, which is not true in the threat model of this work, see Section~\ref{sub:threat_model}. 
To address this, \textit{local DP} (LDP)~\cite{truex2020ldp} was proposed, where each \user adds noise to its local update before sending it out for aggregation, in a way that ensures the \user's own dataset is protected from extraction. Unfortunately, LDP generally results in degraded model utility due to the addition of large amounts of noise to every \user's update~\cite{wei2020federated}. %
\textit{Distributed DP (DDP)} was proposed as a popular middle ground between CDP and LDP, where multiple \users independently add noise to their update, that is sufficient to ensure their datasets are protected from an extraction adversary, but only as long as their updates are aggregated before the adversary observes them. Through combination with \SA~\cite{kairouz2021distributed,agarwal2021skellam,chen2022fundamental} or similar aggregation methods~\cite{bittau2017prochlo}, where the \cp can only view aggregated updates, DDP ostensibly ensures that the \cp cannot extract individual data points. But, importantly, this assumes that a large fraction of \users participating in the FL round are honest and add their share of the noise. 
We discuss the applicability of this assumption in the real world in \Cref{sub:adv_model}.

\section{Threat Model and Assumptions}
\label{sub:threat_model}

We characterize our threat model and assumptions in terms of our adversary and the considered FL setup.

\subsection{Adversary}
\label{sub:adversary}
Our adversary is the \cp, and their goal is to infer individual \users' local sensitive data points. Note that the background in \Cref{sec:background} implies that the \cp can---whenever they choose to---(1) control the weights of the shared model, (2) select which of the $\totalusers$ \users participate in each round, and (3) provision new \users into the pool (including \textit{sybils} controlled by the \cp).\footnote{Capabilities (2) and (3) have been demonstrated in the real world as Google researchers introduced 189 sybils devices into the Gboard FL system and made them participate in the protocol along with real \users~\cite{ramaswamy2020training}.} We assume that the total fraction of sybils out of the $\totalusers$ \users is small.

We, furthermore, assume the \cp is \emph{occasionally malicious} (OM), meaning they behave maliciously in only a few rounds of the protocol. 
When this happens, they can exert the above capabilities (1-3) adversarially. 
Do note that the \cp here does not deviate from the protocol and restricts themselves to only use valid operations (1-3) in the FL protocol. 
An OM \cp ensures the attack remains stealthy, and also allows the \cp to train a model that has high utility over the non-malicious rounds, which is an expected product of FL.

\subsection{Adversarial Model}
\label{sub:adv_model}
By assuming a server's ability of introducing hundreds of sybil devices into the pool of active users, this work relies on a strong adversarial model. 
Yet, industry actors that deploy FL protocols in real-world applications have shown that this scenario is a realistic threat~\cite{ramaswamy2020training}.
For the purpose of a research project, they successfully introduced hundreds of sybil devices in the real-world FL training of the Google keyboard and let them train along the real users for some time to manipulate the model training.
Since obtaining and introducing sybil devices does not only require configuration but also a non-negligible financial overhead, we expect %
this type of manipulation to be reserved to adversaries that can afford it.

Another setting in which we think our attack is practical is the one of colluding employees. %
Since prior research has documented that it generally takes a small number (e.g., 2-3) of employees to approve changes to a company's code base~\cite{potvin2016google,sadowski2018modern}, we could imagine that these employees would collude. Jointly, these employees could then maliciously exploit the servers' abilities of user sampling and controlling the shared model. %
We account for this scenario by our OM threat model, and argue that by preventing this type of attack, a company can protect itself against actions that might harm their reputation.

\subsection{FL Setup}
\label{sub:fl_setup}

Our departure point are FL protocols deployed in real-world applications, such as the one described in~\cite{bonawitz2019towards}. These protocols initially focused on data minimization only. 
We extend them with \SA and DDP, two defenses dedicated to additionally providing privacy protection for the FL protocol.
We note that we chose to study an instantiation of FL with \SA and DDP because it is the combination of published techniques that holds the strongest promise in the presence of an untrusted \cp.\footnote{Indeed, the key alternative to FL with \SA and DDP would be FL with LDP (local DP guarantees) but this alternative is not appealing because it comes at a significant utility cost for the \cp.}
Note, however, that FL with \SA and DDP is not as widely deployed as vanilla FL is.
This is mainly due to some increased communication costs~\cite{bell2020secure} and the computational overhead of adapted mechanisms~\cite{agarwal2021skellam, kairouz2021distributed}.

\section{Attacking FL under \SA and DDP}
\label{sec:fl+SA+DP}

In this section, we study FL extended by DDP and \SA---considered as a strongly privacy-protective instantiation of the protocol---and show that the \cp can still reconstruct sensitive information about the \users' training data.
We also forge an intuition of what factors contribute most to the leakage.
Based on our findings, in \Cref{sec:recommendations}, we discuss future research and implementation towards private FL. %

For successful data reconstruction under DDP and \SA, the \cp has to make use of the following three capabilities which it naturally holds in FL:
\begin{enumerate}
    \item \emph{Introducing sybil devices:} The \cp needs to be able to introduce a fraction of manipulated devices in the FL protocol (see \Cref{sub:adv_model}). These devices can return arbitrary gradients, chosen by the \cp. In particular, they can contribute zero gradients to the \SA. 
    \item \emph{Controlling the \user sampling:} To ensure that the sybil devices are sampled for \SA together with a target \user, the \cp needs to control the \user sampling.
    \item \emph{Manipulating the model weights:} For improved data reconstruction performance, the \cp can manipulate the shared model's weights, for example, relying on one of the methods discussed in \Cref{sec:attacking_vanilla_FL}.
\end{enumerate}

While the first two capabilities enable the \cp to circumvent \SA and to leave the gradients of a target \user with insufficient amount of noise for privacy protection under DDP, the third capability increases the amount of individual training data that can be reconstructed and extends the attack to other model architecture types.

\myparagraph{Attack flow.}
\label{sub:sa+ddp_attack_flow}
Our attack aims at reconstructing the private data of one target \user per malicious round in the FL training. 
To do so, the attack needs to  circumvent the \SA (\Cref{sub:circumvent_sa}), then exploit the weak privacy guarantees of DDP from a user's perspective (\Cref{sub:exploit_ddp}), and finally reconstruct the target \user's individual training data points (\Cref{sub:ddp_reconstruction}). See \Cref{fig:FL+SA+DDP} for the course of our attack.

\subsection{Circumventing \SA}
\label{sub:circumvent_sa}
In our attack, the \cp circumvents \SA by sampling the target \user together with maliciously controlled sybil devices for the given training round.
Since for each round, $\selectusers$ \users are sampled for participation, the \cp needs to control at least $\selectusers-1$ sybil devices.
It has been shown in previous work~\cite{ramaswamy2020training} that inserting an arbitrary number of sybil devices into real-world FL deployments is practically feasible as we discuss in \Cref{sub:adv_model}.

Since \SA-protocols provide their guarantees under the assumption that a certain fraction of \users is honest~\cite{bonawitz2017practical,bell2020secure}, it follows naturally that in the presence of the sybil devices, no guarantees can be provided to the target \user.
This is because when the gradients are aggregated over multiple \users, and all but the target \user contribute arbitrary gradients, known to the \cp, the \cp can extract the target \user's gradients perfectly.
In the easiest case, the sybil devices contribute all zero-gradients, such that the final aggregate directly only contains the target \user's gradients.

Note that we do not claim that the SA or any of the underlying cryptographic primitives are broken.
We solely observe that \SA relies on the assumption that the clients participating in the execution are real clients and not maliciously controlled by the server~\cite{bell2020secure}.
However, in FL with an untrusted server, the \users cannot verify this assumption.
We show through our attack that this has severe implications on their privacy guarantees.

Pasquini~\etal~\cite{pasquini2021eluding} describe a different way to circumvent \SA based on the \cp sending out different models to different \users.
While the models for non-target \users produce zero-gradients, the target \user's model produces non-zero gradients which can be exploited for data reconstruction.
An advantage of this method is that it does not require the \cp to control the \user sampling, or to manipulate a fraction of \users.
Note however that in their scenario, DDP can still be efficiently applied if every \user adds some noise to their (potentially zero) gradients.
As a consequence, the total amount of noise can be sufficient to protect the gradients of the target \user. 
Therefore, in our attack, we rely on the controlled sybil devices to circumvent the \SA.

Note that also alternative mechanisms to implement DDP, such as shuffling~\cite{bittau2017prochlo,erlingsson2019amplification} which can be put into place instead of \SA, can be circumvented by our approach of inserting sybil devices with server-controlled gradients into the protocol.

\subsection{Exploiting DDP Guarantees}
\label{sub:exploit_ddp}

\begin{figure*}[tb]
\centering
\includegraphics[width=0.9\textwidth, trim={2.cm 5.5cm 0cm 0cm},clip]{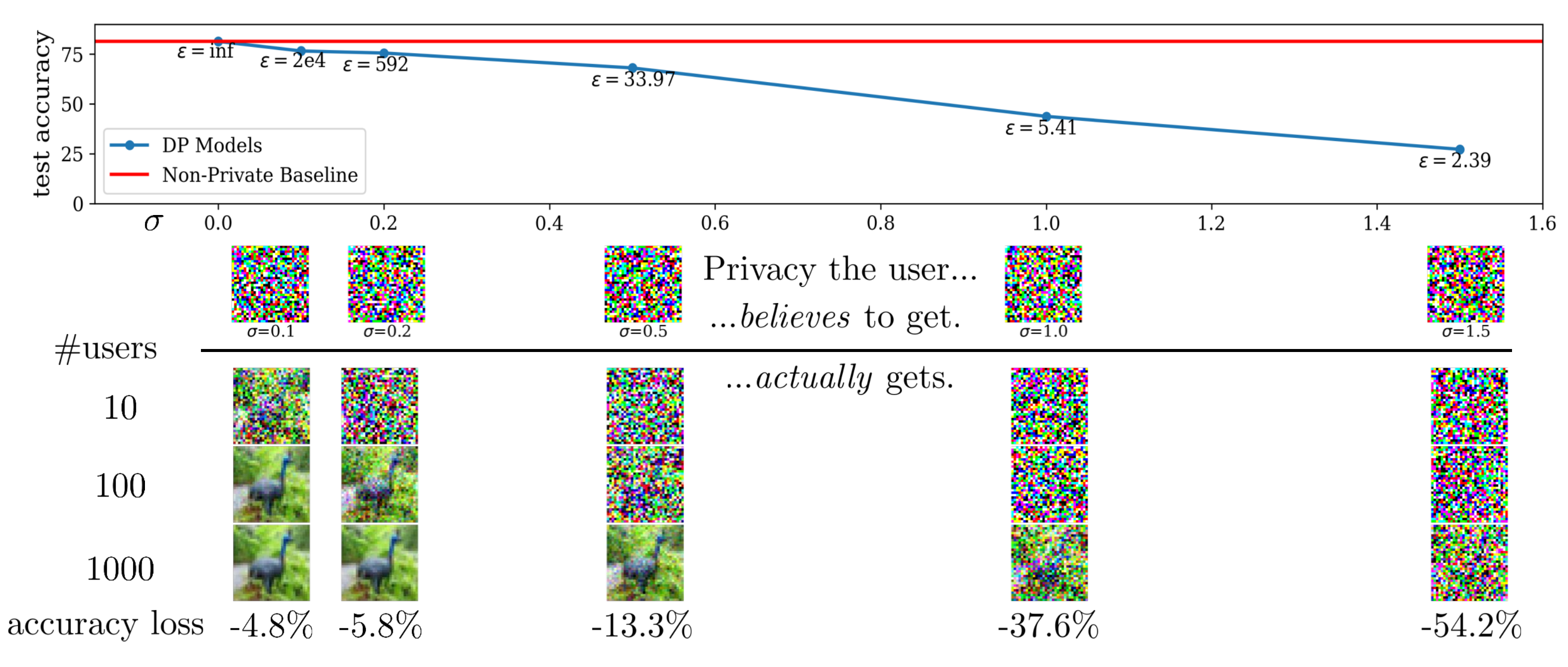}
\caption{
\textbf{Privacy vs. utility trade-offs under DDP/LDP.} Each point on the \textcolor{blue}{blue line} corresponds to a model trained on CIFAR10 with a clipping parameter $c=1$, and the total noise scale $\sigma$ indicated on the x-axis. 
Training was conducted over 100 epochs, the resulting privacy guarantees $\epsilon$ are reported. 
For the non-private baseline $\epsilon=\infty$ (\textcolor{red}{red line}).
We report accuracy loss with respect to this non-private baseline.
The images depicted below the line plot display the rescaled noisy gradients of one data point of an individual \user with noise calculated according to \Cref{eg:ddp} as a function of $\sigma$, $c$, and $\selectusers$ in the training round.
The more \users participate, the less noise every \user needs to add because, during aggregation, the total noise is determined by the sum of the individual noises.
If, however, other \users do not add noise, the locally added noise is the only privacy protection every individual \user has (DDP reduces to LDP with weak privacy guarantees). 
As a consequence, there is a discrepancy between the privacy the \user believes to get and the privacy they actually get:
The images above the black line visualize the \user's belief on what the \cp can extract from their gradient, the images below the black line visualize what the \cp can actually extract under our attack.}
\label{fig:DP-utility}
\end{figure*}

If DDP is in place, the gradients of the target \user will be slightly noisy---even with successful circumvention of \SA.
However, by design of DDP, the amount of noise added by each user is typically insufficient to provide a meaningful privacy guarantee \textit{from the user's perspective}~\cite{kairouz2021distributed}. By meaningful privacy guarantees we mean, guarantees equivalent to what one would obtain in the LDP definition. 
This is in fact how DDP obtains a utility gain over LDP, which would have inserted sufficient noise to obtain per-user privacy guarantees that are independent of other users: 
DDP assumes all users will add enough noise so that the \emph{aggregate} is sufficiently noised whereas LDP assumes each user adds enough noise to obtain \emph{privacy in isolation}. 
As a consequence, in DDP, each \user can add less noise locally than required for the desired total privacy level, resulting in more utility.
In contrast, the guarantee provided by LDP allows the user to not trust the \cp or other users.

Concretely, in an LDP version of FL, the noise added by each \user depends solely on the noise scale $\sigma$ and the clipping parameter $c$ of the application.
As a consequence, the local noise is sampled from a Gaussian distribution according to
\begin{align}
    \mathcal{N}(0, \sigma^2c^2)\text{.}
\end{align}

In contrast, in DDP, the amount of noise added by each individual \user additionally takes the number of \users who participate in the round into account~\cite{truex2019hybrid}.
Assuming that $\selectusers$ \users are sampled for participation, this results in a local addition of Gaussian noise sampled from
\begin{align}
\label{eg:ddp}
    \mathcal{N}\left(0,  \frac{\sigma^2}{\selectusers-1}c^2\right)\text{~\cite{truex2019hybrid}.}
\end{align}

In \Cref{fig:DP-utility}, we present the privacy-utility trade-offs resulting from training models on the CIFAR10~\cite{krizhevsky2009learning} dataset as a function of the total noise scale $\sigma$ and the resulting models' accuracy on a test set. 
We train the private models with a state-of-the-art framework for DP-training\footnote{\url{https://github.com/ftramer/Handcrafted-DP}. Note, however, that our reported accuracy and achieved privacy levels $\epsilon$ cannot directly be compared with the values reported in the repository. This is because we use different noise scales than they do and train the model for $100$ epochs while they only train for $30$ epochs.} in which all hyperparameters and model architecture are tuned for the~task. %

\Cref{fig:DP-utility} provides two main insights. 
First, unsurprisingly, given the privacy-utility trade-offs mentioned above, the model utility decreases when the total noise scale $\sigma$ increases.
Second, the figure shows that the more \users participate in a given training round, the less noise each \user needs to add locally.
This results from \Cref{eg:ddp} which relies on the total noise being aggregated over all participating \users before sharing the aggregated gradients with the \cp.

DDP assumes that each \user is \emph{honest} and adds the required noise to their gradients.
However, if even one of the \users adds less than the amount of noise it should add, the desired total privacy guarantees cannot be reached. 
Even worse, if, as described in \Cref{sub:sa+ddp_attack_flow}, a target \user in FL is sampled for participation solely with controlled sybil devices that do not provide any noise for aggregation, the local noise added by the target \user represents the only protection for its gradients. 

These results mean that there is a tension between (1) the guarantee claimed by the \cp (and other \users) in DDP and (2) the guarantee that a user who does not trust this \cp can rely on. This will lead the \cp optimizing for model utility to request that users add less noise to their gradients than what is needed for individual users to protect their data from leaking to an untrusted \cp.

\subsection{Reconstructing Data}
\label{sub:ddp_reconstruction}

In \Cref{sec:attacking_vanilla_FL}, we presented different attacks that rely on manipulations of the shared model to extract individual \users' training data points.
In principle, each of these attacks can be included to perform data reconstruction in our FL+SA+DDP setup.
However, the attack by \cite{wen2022fishing} extracts individual data points over several rounds of the FL protocol.
In our setup, due to the \cp's OM nature, single-round attacks are preferable.
These allow the adversary to stay more inconspicuous and to train a more meaningful shared model throughout the benign rounds.
The attack by \cite{fowl2021robbing} relies on manipulations of the model architecture, which are more detectable than manipulations of model parameters, such as in~\cite{boenisch2021curious}.
Finally, our experimental evaluation highlighted a significant advantage of~\cite{boenisch2021curious} in comparison to~\cite{fowl2021robbing} for data reconstruction under noise.
\cite{boenisch2021curious}'s trap weights yield redundancy in the extracted data, \ie the same data point can be extracted multiple times from gradients of different weight rows.
We thoroughly investigate this effect in \Cref{sub:redundancy}.
The redundancy of extracted noisy data can be exploited to average out the effect of the noise and yield higher-fidelity data reconstruction, as we will show in \Cref{sub:reconstruct_DDP}.
In contrast, due to the nature of their attack, in~\cite{fowl2021robbing}, each data point is only extractable once.

\section{Evaluation of the Attack against SA+DDP}
\label{sec:experiments}

In this section, we present a practical evaluation of our attack against FL protected with SA and DDP.
We first present our experimental setup.
Then, we evaluate direct extraction of noisy gradients under DDP.
We move on to experimentally evaluate the redundancy of extracted data with \cite{boenisch2021curious}'s trap weight approach, and discover how this redundancy can be exploited for higher-fidelity data reconstruction under noise.
We illustrate this with empirical results reconstructing image and text data.

\subsection{Experimental Setup}
\label{sub:setup}
We operate in a cross-device FL setup and perform training on the CIFAR10~\cite{krizhevsky2009learning} dataset.
We evaluate extraction on different mini-batch sizes $B\in\{10, 20, 100\}$.
We split the CIFAR10 training data at random and iid between the \users. 
\cite{boenisch2021curious}~shows that their trap weights' extraction success is equal for iid and non-iid distribution, even for the most extreme scenario where every data point in a given mini-batch stems from same class.
Following~\cite{boenisch2021curious}, we also experiment with the IMDB dataset for sentiment analysis and the  and distribute data the same way.
An extended experimental evaluation on two additional image datasets (MNIST and ImageNet), and two textual spam classification datasets (Spam Mails and SMS Spam Collection) can be found in Appendix~\ref{app:experiments}.

To evaluate different setups for DDP, we select $\{10, 100, 1000\}$ \users for participation in a given round of the FL protocol.
To circumvent the \SA, as described in the previous section, we sample one target \user together with sybil devices which all return zero gradients.
Other than that, we follow~\cite{boenisch2021curious}'s experimental setup, use their six-layer fully-connected neural network and embedding architecture (their Table 7 and 8), initialize the first fully-connected layer with their trap weights for individual data point extractability.
To project received gradients of the first fully-connected layer's weight matrix back to their input domain, we rely on~\cite{boenisch2021curious}'s Equation (5) which shows that it is sufficient to rescale the gradient of the weights with the inverse of the gradient of the bias for perfect data extraction.
Note that in our case, due to DDP, noise is added not only to the gradient of the weights but also to the gradient of the bias.
Therefore, not only the extracted gradients but also our scaling factor are noisy, resulting in the rescaled gradients not being a perfect reconstruction of the original input data.

We report the DDP-setup per-round through three parameters required to determine the noise magnitude according to \Cref{eg:ddp}, namely the DP clip norm $c$, the DP noise multiplier $\sigma$, and the number of selected \users in this round $\selectusers$. This enables us to understand how sensitive the attack is to choices for these hyperparameters without making any assumptions about other hyperparameters of the training run (e.g., the number of training steps).

\subsection{Noisy Data Extraction}
\label{sub:noisy_extraction}

\begin{figure}[t]
    \centering
    \includegraphics[width=\linewidth]{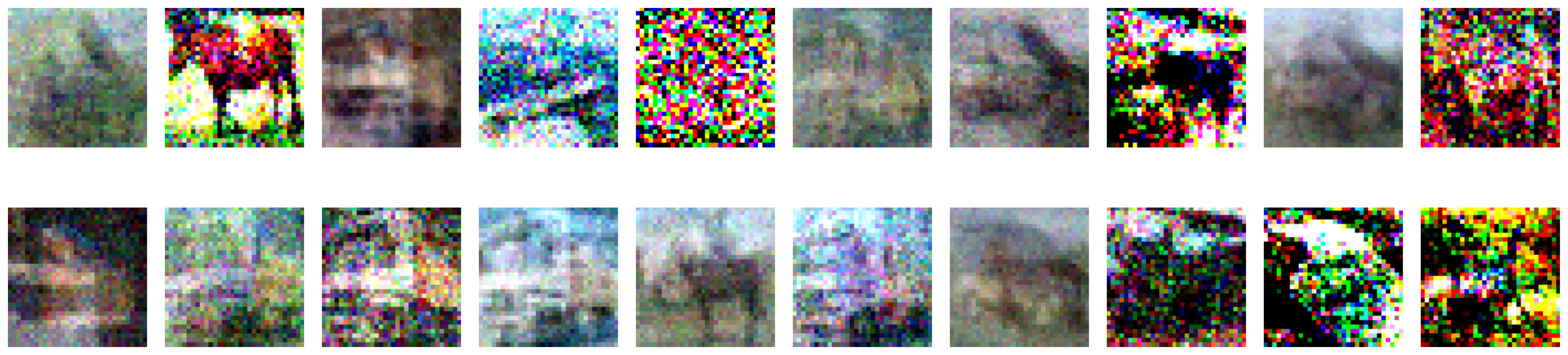}
    \caption{\textbf{Directly Extracted Data under DDP.} Rescaled clipped and noised gradients from a mini-batch with 20 data points from CIFAR10 dataset. 
    DDP setup: $c=1$, $\sigma=0.1$, and $\selectusers=100$.}
    \label{fig:dp-before-cluster}
\end{figure}

We first perform direct extraction from noisy gradients.
The extraction of data points works exactly the same way as for vanilla FL~\cite{boenisch2021curious}, by initializing the model with trap weights before sending it to the target user, and then projecting their received gradients back to the input domain.
\Cref{fig:dp-before-cluster} shows the full resulting extracted data in a setup with a mini-batch of 20 data points, $c=1$, $\sigma=0.1$, and $\selectusers=100$.
While some noisy reconstructed data points resemble the original training data, other are less recognizable because they are an overlay of multiple data points, or are dominated by the added noise. DDP setup: $c=1$, $\sigma=0.1$.
We report further results for MNIST and ImageNet in Appendix~\ref{app:exp_img}.

\begin{table}[tb]
    \centering
    \begin{tabular}{ccc}
    \toprule
    \textbf{Number of} & \% \textbf{Individually} & \textbf{Reconstruction} \\
    \textbf{Benign \Users} & \textbf{Reconstructable Data} & \textbf{SNRs} \\
        \midrule
1  &   0.867 & 0.015 \\
2  &   0.800 & 0.012 \\
3  &   0.733 & 0.011 \\
4  &    0.717 & 0.010 \\
5  &   0.633 & 0.010 \\
10 &    0.400 & 0.008 \\
15 &     0.317 & 0.008 \\
20 &     0.250 & 0.008 \\
30 &     0.133 & 0.007 \\
40 &     0.000 & 0.007 \\
50 &     0.000 & 0.007 \\
         \bottomrule
    \end{tabular}
    \caption{\textbf{Influence of Fraction of Sybil Devices.} Results for FL hardened by SSA and DDP with 50 participants, each holding $B=20$ data points. We replaced a varying fraction of \users by sybil devices and measured number of individually extractable data points from the target \user and average SNRs over all their reconstructions. The more benign \users participate in the round, the less effective data reconstruction becomes.}
    \label{tab:influence_of_proportion_of_sybils}
\end{table}

\begin{figure}[t]
    \centering
    \includegraphics[width=\linewidth]{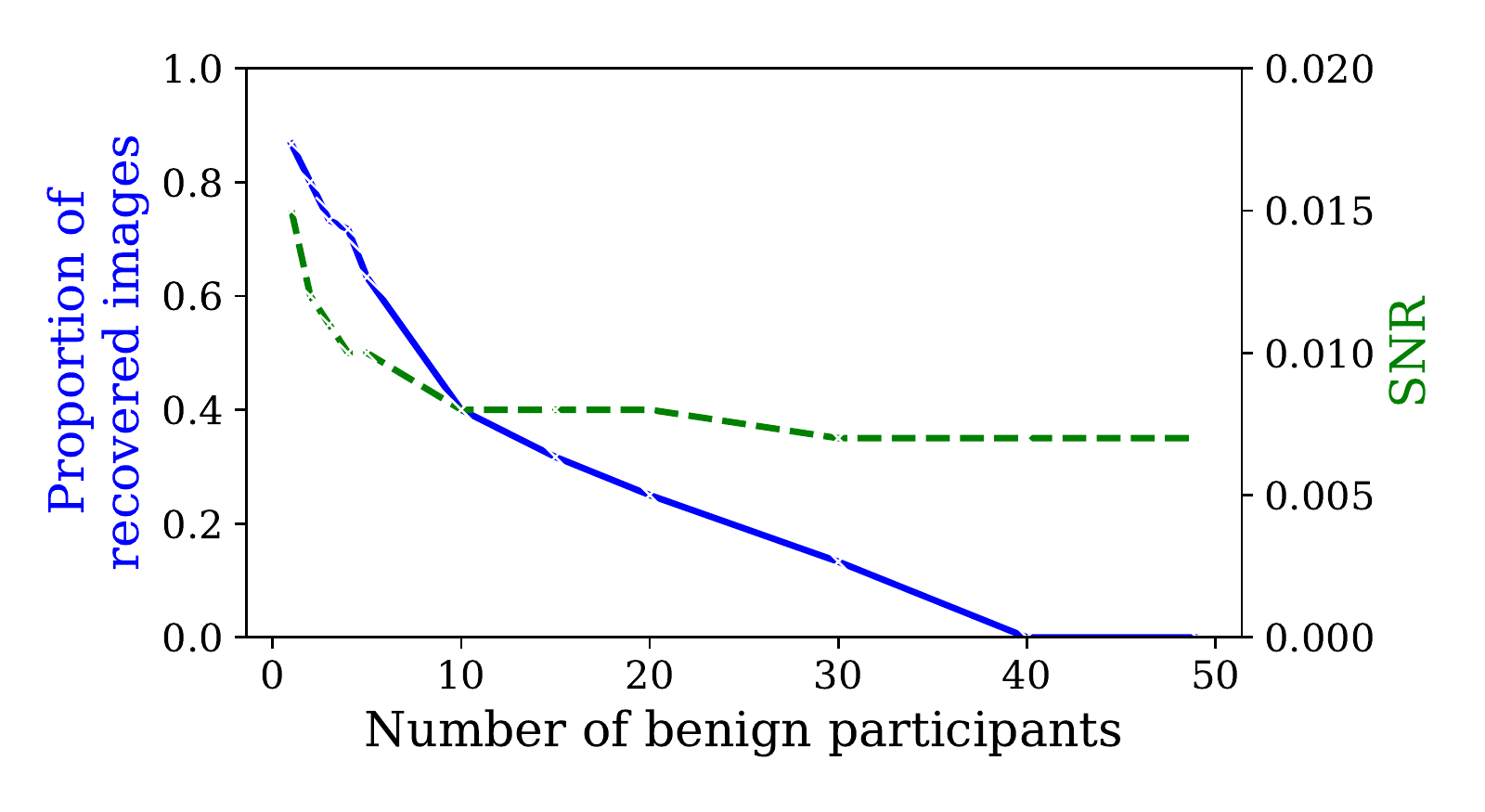}
    \caption{\textbf{Influence of Fraction of Sybil Devices.} Visual depiction of the results presented in \Cref{tab:influence_of_proportion_of_sybils} for FL hardened by SSA and DDP with 50 participants, each holding $B=20$ data points.}
    \label{fig:influence_of_proportion_of_sybils}
\end{figure}

\myparagraph{Effect of Fraction of Sybils.} 
We, furthermore, conducted experiments to quantify the effect of not replacing all other $\selectusers-1$ \users by sybil devices, but only a fraction of the other \users.
Therefore, we conducted experiments with extracting data from one round of the FL protocol with 50 participants, each holding a mini-batch of 20 data points.
Our goal was to study what privacy gain the presence of other benign \users incurs on the target \user.
We visualize our results in \Cref{tab:influence_of_proportion_of_sybils} and \Cref{fig:influence_of_proportion_of_sybils}.
They suggest that when the target \user is sampled solely with sybil devices (row 1), the \cp is able to extract 95\% of their individual training data points individually, protected solely by the noise added locally according to DDP.
The more benign \users participate in the protocol round, the lower SNRs of the reconstructed data from the target \user, and the fewer of the target \user's individual data points can be individually extracted.
This effect stems from the aggregation within the \SA, which overlays gradients from all \users before sharing them with the \cp.
Our results are aligned with findings by \cite{boenisch2021curious} who showed that averaging over several mini-batches (which is precisely the effect of the \SA) degrades extraction success.

\myparagraph{Sufficiently Protective Noise.} 
Deciding at which point, \ie, under the influence of how much noise, the reconstruction of a data point is sufficiently close to the original data point is orthogonal to this work.
In particular, it will depend on the specific domain, task, and \user-preference.
However, \users in FL should assume that the \cp can extract individual data points such as the ones depicted in \Cref{fig:dp-before-cluster} from their gradients.

In the following, we will show how the \cp can improve the fidelity of extraction by leveraging the redundancy of extractable data due to the trap weights.

\subsection{Redundancy in Extracted Data}
\label{sub:redundancy}

This section studies redundancy in extracted data of the trap weights method and their effect on the fidelity of reconstructed data.

\begin{figure}[tb]
\centering
\begin{subfigure}[b]{0.23\textwidth}
\centering
\includegraphics[width=\linewidth]{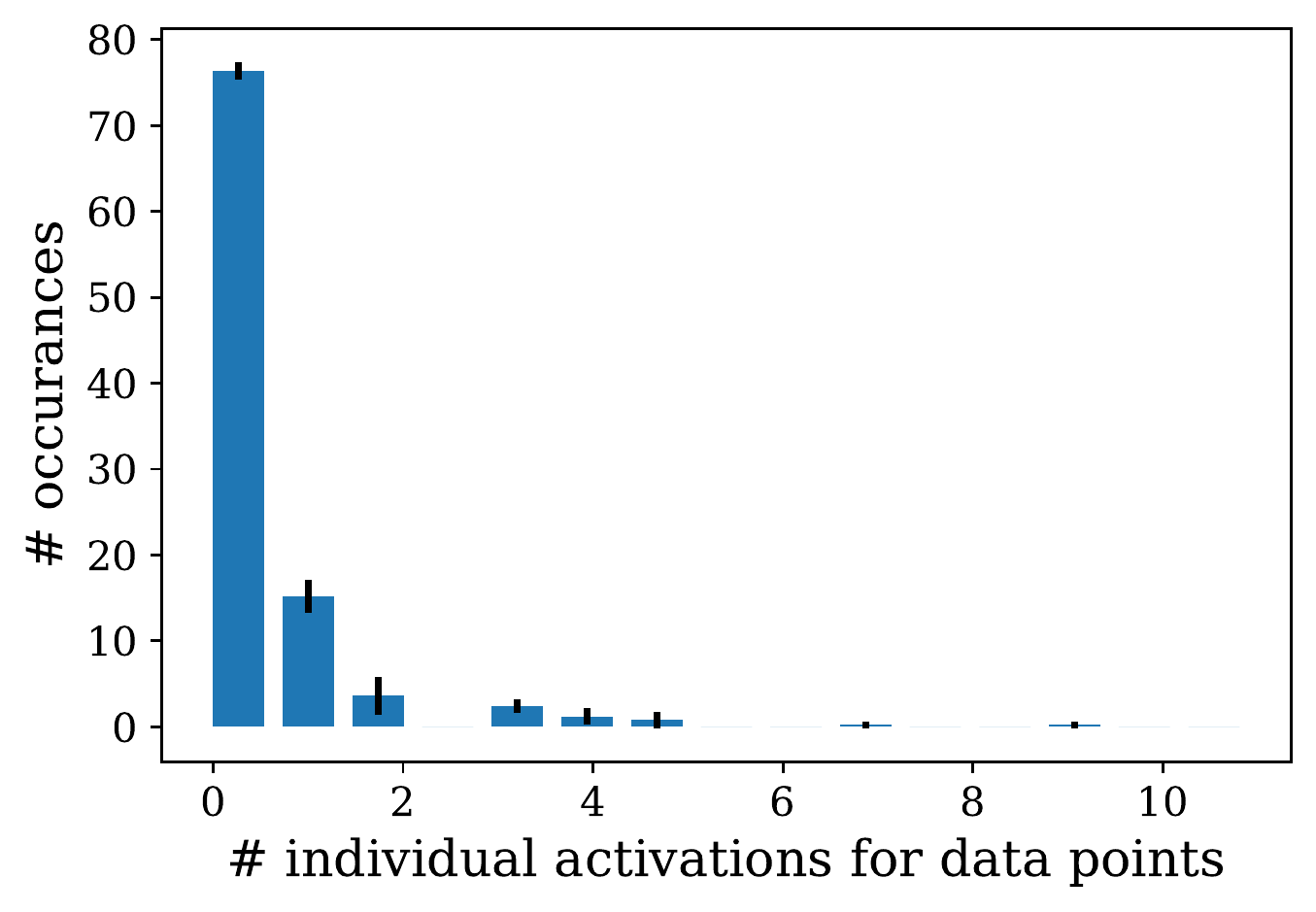}
\caption{Random weights.}
\end{subfigure}     
\begin{subfigure}[b]{0.23\textwidth}
\centering
\includegraphics[width=\linewidth]{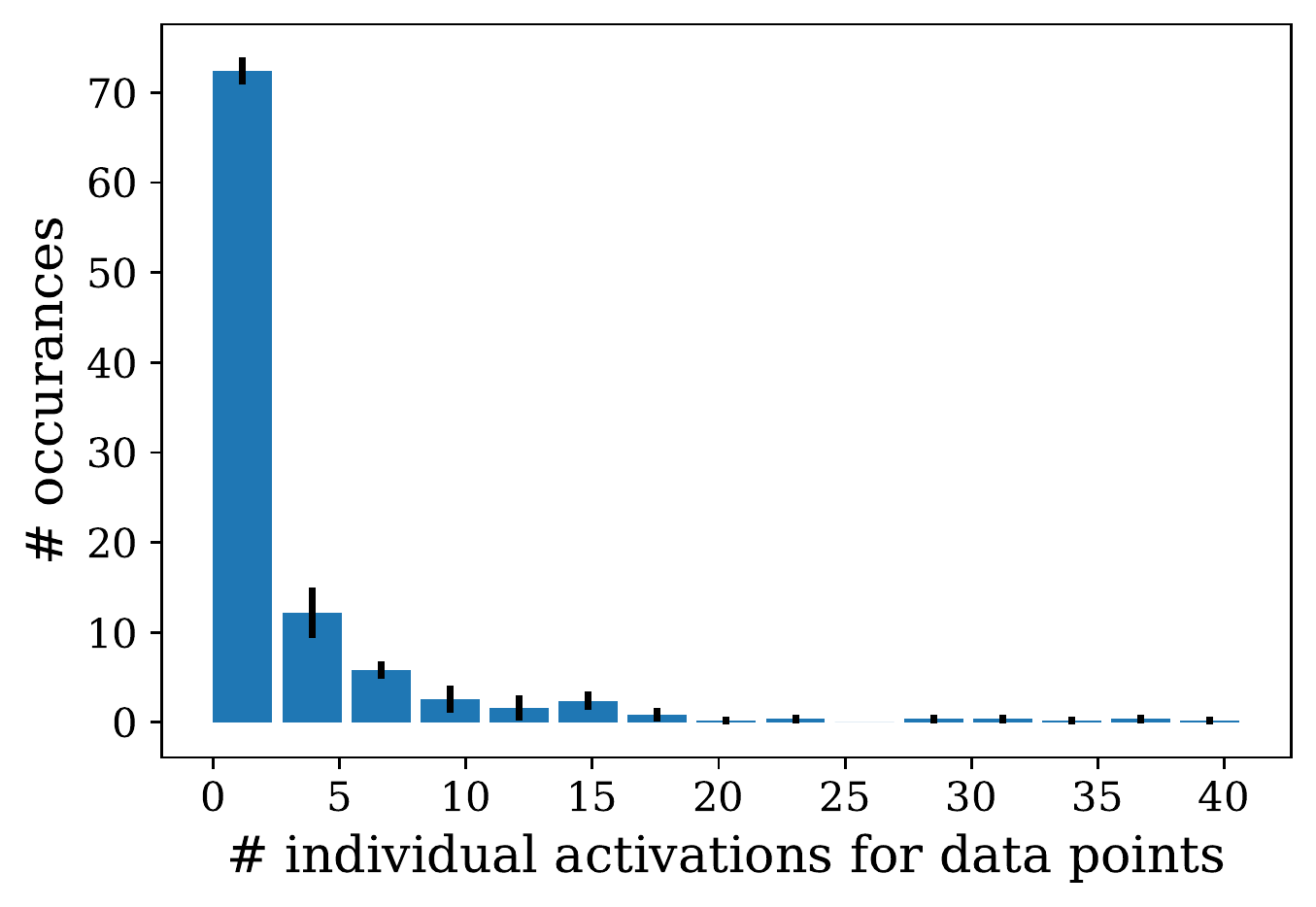}
\caption{Trap weights.}
\end{subfigure}   
\caption{\textbf{Number of Activations per Data Point.} The number of times each of the 100 data points is individually extractable from the model gradients. 
The same data points are individually extractable from many more different gradients when using trap weights which enable us to use redundancy for better data reconstruction. 
Results are averaged over five different random and trap weight model initializations.
}
\label{fig:extractability_q1}
\end{figure}

\myparagraph{Redundancy in Extractable Data Points.}
\label{sub:extractability}
We first study direct redundancy by analyzing \emph{how often} each data point in a mini-batch with $B=100$ is individually extractable from the rescaled gradients.
The results depicted in~\Cref{fig:extractability_q1} suggest that the trap weights, first of all, make more data points individually extractable in contrast to random model initializations, but also cause the same data points to be individually extractable from many more different weight rows' gradients (up to 70 times over the 1000 neurons and their respective weight rows).

\begin{figure}[tb]
\centering
\begin{subfigure}[b]{0.23\textwidth}
\centering
\includegraphics[width=\linewidth]{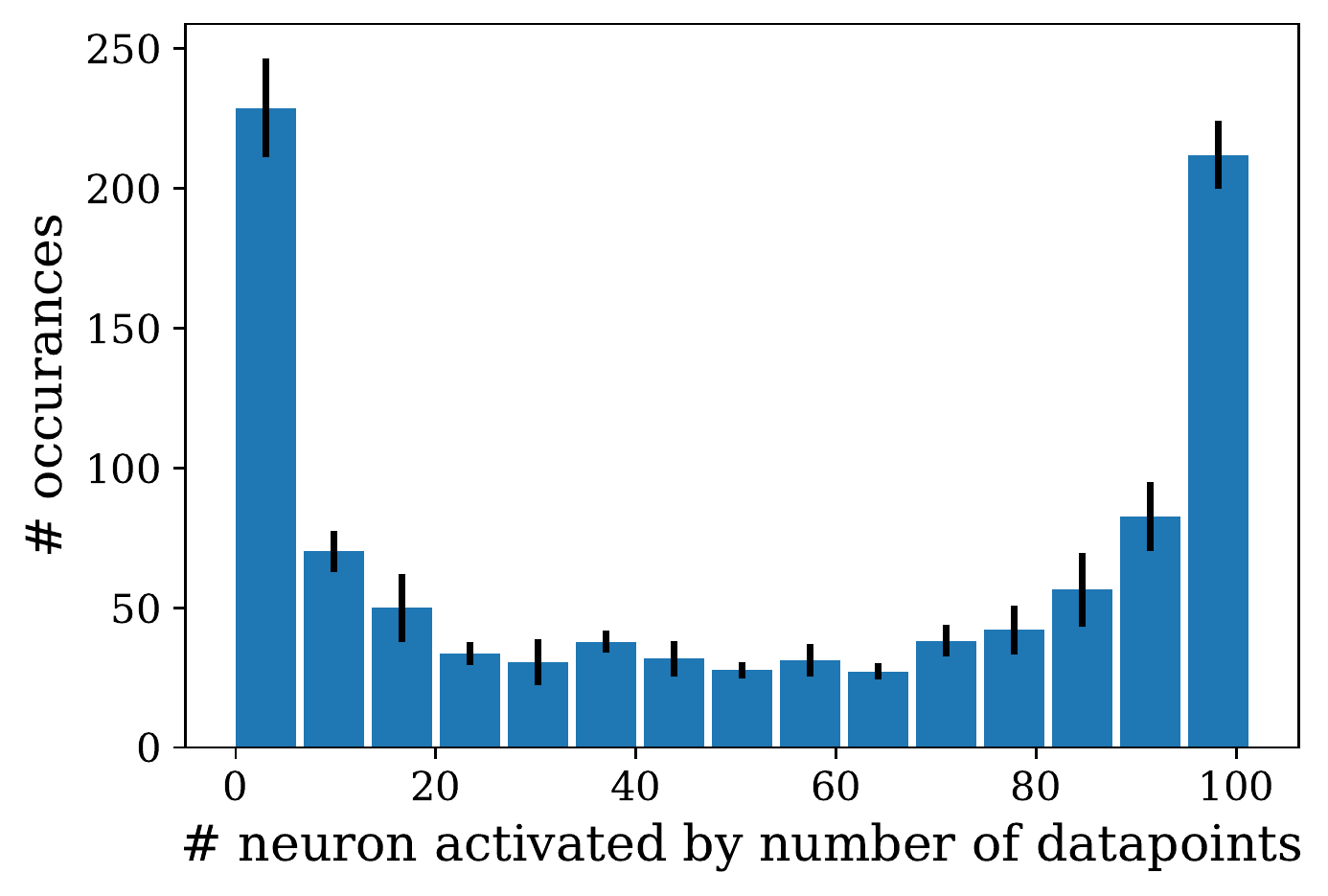}
\caption{Random weights.}
\end{subfigure}     
\begin{subfigure}[b]{0.23\textwidth}
\centering
\includegraphics[width=\linewidth]{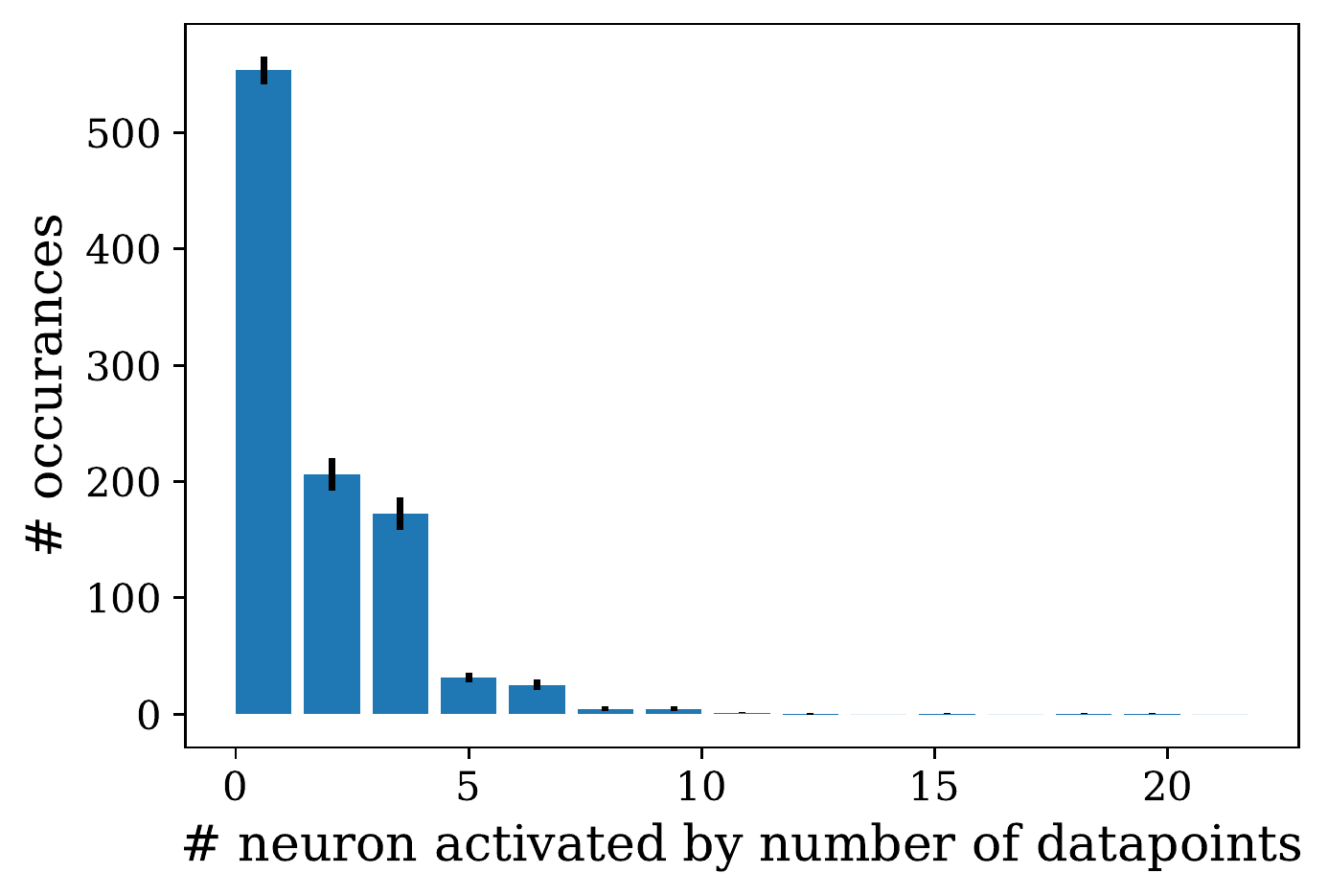}
\caption{Trap weights.}
\end{subfigure}   
\caption{\textbf{Number of Activations per Neuron.} Number of data points that activate each one of the 1000 neurons. 
Individual neurons are activated by fewer data points (less overlay) when using trap weights which enable better data reconstruction. 
Results are averaged over five different random and trap weight model initializations.
}
\label{fig:extractability_q3}
\end{figure}

\myparagraph{Sparsity in Extractability.}
\label{sub:effectiveness_trap_weights}
We, furthermore, evaluate \emph{by how many data points} each neuron gets activated.
This is equivalent to the question how many data points cause a positive input to each neuron.
\Cref{fig:extractability_q3} highlights that with randomly initialized weights, neurons are activated by many more data points than with the trap weights, which causes that many data points overlay in a single gradient and we cannot extract them individually.

\begin{figure}[t]
\centering
\centering
\includegraphics[width=0.65\linewidth]{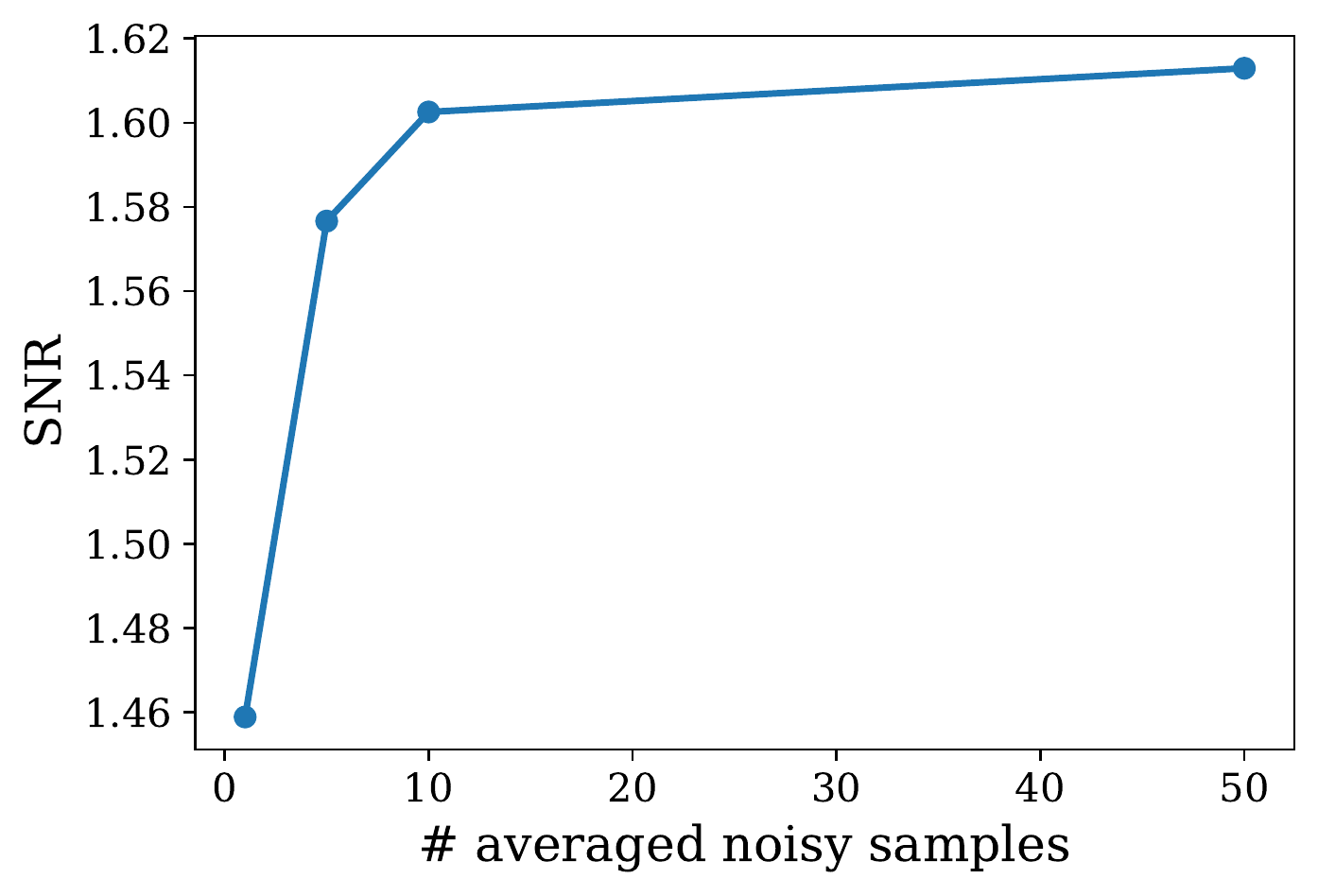}
\includegraphics[width=0.65\linewidth]{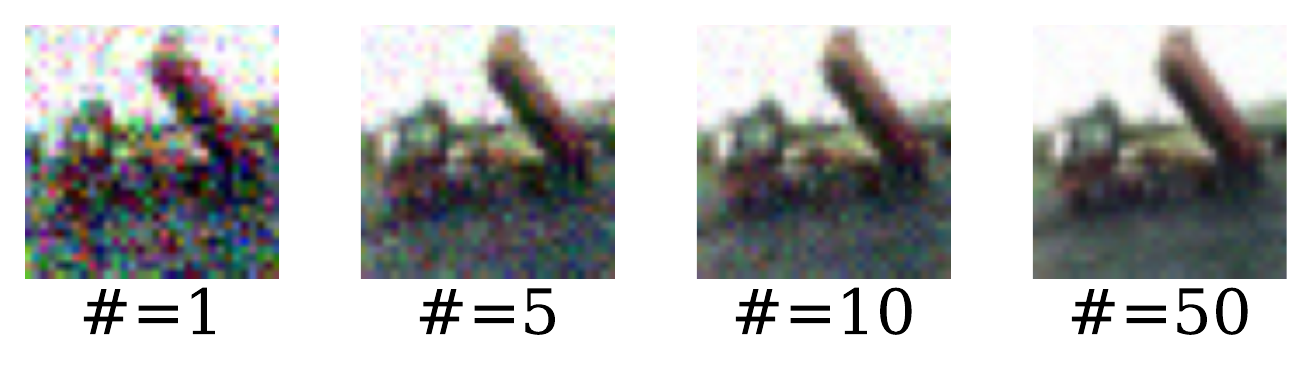}
\caption{\textbf{Averaging out Noise.} Mean value over \#-many noisy reconstructions of the same data point (bottom); corresponding mean image's SNR (top). 
DDP setup: $c=1$, $\sigma=0.1$, and $\selectusers=100$.
Over an increasing number of reconstructions, the local noise averages out, yielding higher-fidelity images and increased SNR.}
\label{fig:SNR_DP}
\end{figure}

To improve fidelity of reconstruction, we can leverage both the redundancy of extractable data and the sparsity in the extracted data.
By averaging redundant noisy data points, the signal-to-noise ratio (SNR) of reconstructed data increases as noise averages out.
We visualize this effect in \Cref{fig:SNR_DP}.
Also sparsity can be exploited.
The fewer data points activate a neuron, the fewer data points contained in the overlay of the rescaled gradient.
Hence, each individual data point's signal is more clearly present and identifiable in the rescaled gradients. 
In the following section, we will show how this can be used to yield higher-fidelity reconstruction in the image domain through clustering.

\subsection{Improving Noisy Data Reconstruction}
\label{sub:reconstruct_DDP}

The previous sections highlight that DDP reduces to LDP with weak privacy guarantees from an individual \user's perspective when other \users are untrusted with their noise addition.
In this section, we show how we can even improve data reconstruction in this setup, further amplifying the small signal in the extracted gradients.
We evaluate improvements for data reconstruction from noisy gradients computed under DDP on \textit{image and textual data}.
All improvements solely rely on post-processing steps to reduce the effect of the noise.

\myparagraph{Image Data.}
Due to the local clipping and noise addition by the \users, the data points extracted from the gradients are not perfect reconstructions of the original data points.
We can still improve reconstruction quality by leveraging redundancy and sparsity in the gradients to average out the added noise, as highlighted in the previous section.
However, without knowledge of the \users data, the \cp has no means of determining which data points activate which neurons a priori.
Therefore, it is unclear which rescaled gradients need to be averaged to improve reconstruction fidelity.

\begin{figure*}[t]
    \centering
    \includegraphics[width=0.8\linewidth, trim={1.5cm 4.5cm 1.5cm 0cm},clip]{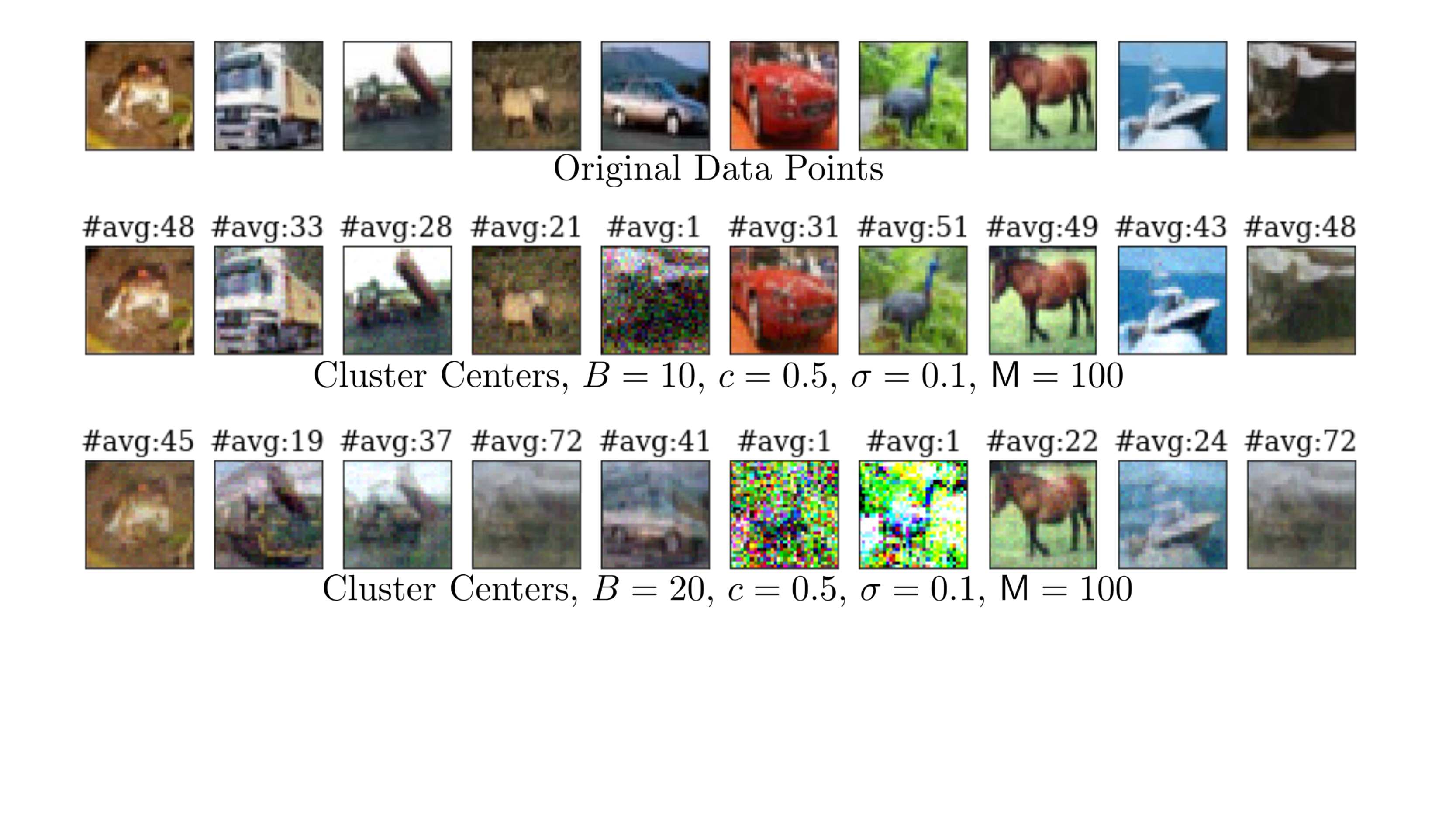}
    \caption{\textbf{Similarity Clustering} to improve noisy data extraction. Original data points and average clusters obtained from the rescaled gradients depicted in \Cref{fig:dp-before-cluster}. First 10 original training data points from the CIFAR10 dataset (top row).
    Averaged clusters of 10 data points reconstructed from the gradients for mini-batch size $B=10$ (mid row), and $B=20$ with the first 10 examples depicted (bottom row).
    The numbers above the images indicate how many noisy reconstructions were averaged to obtain that image.}
    \label{fig:cluster}
\end{figure*}

To overcome this limitation, we employ \textit{similarity clustering}.
In this approach, the \cp first filters out extracted data points with a SNR below~$1$.
This prevents too noisy instances from degrading performance.
In the following \Cref{sub:different_gradient_magnitudes}, we will discuss why different extracted data points have different SNRs.
Then, the \cp runs a simple~$k$-means clustering on the extracted data, and finally averages all per-cluster data points.
Thereby, we do not only leverage redundancy in individually extracted data points, but also the sparsity.
The signal from gradients that represent an overlay of very few data points can meaningfully contribute to the improved signal.
We evaluate this approach with different noise scales and mini-batch sizes~$B$.
Note that the number~$k$ of clusters has to be chosen in accordance with the mini-batch size if we want to be able to reconstruct every data point.
Our evaluation suggests that clustering works best when~$k \ge 2 B$.

In \Cref{fig:cluster}, we depict the results of our clustering on data points from the CIFAR10 dataset with a DDP setup with $c=1$, $\sigma=0.1$ and $\selectusers=100$.
The top row depicts 10 original data points, the mid and bottom rows show the closest averaged clusters for mini-batches of size 10, and 20 respectively.
The more instances are available for averaging, the better the resulting per-cluster averages.

\begin{figure}[t]
    \centering
    \includegraphics[width=\linewidth]{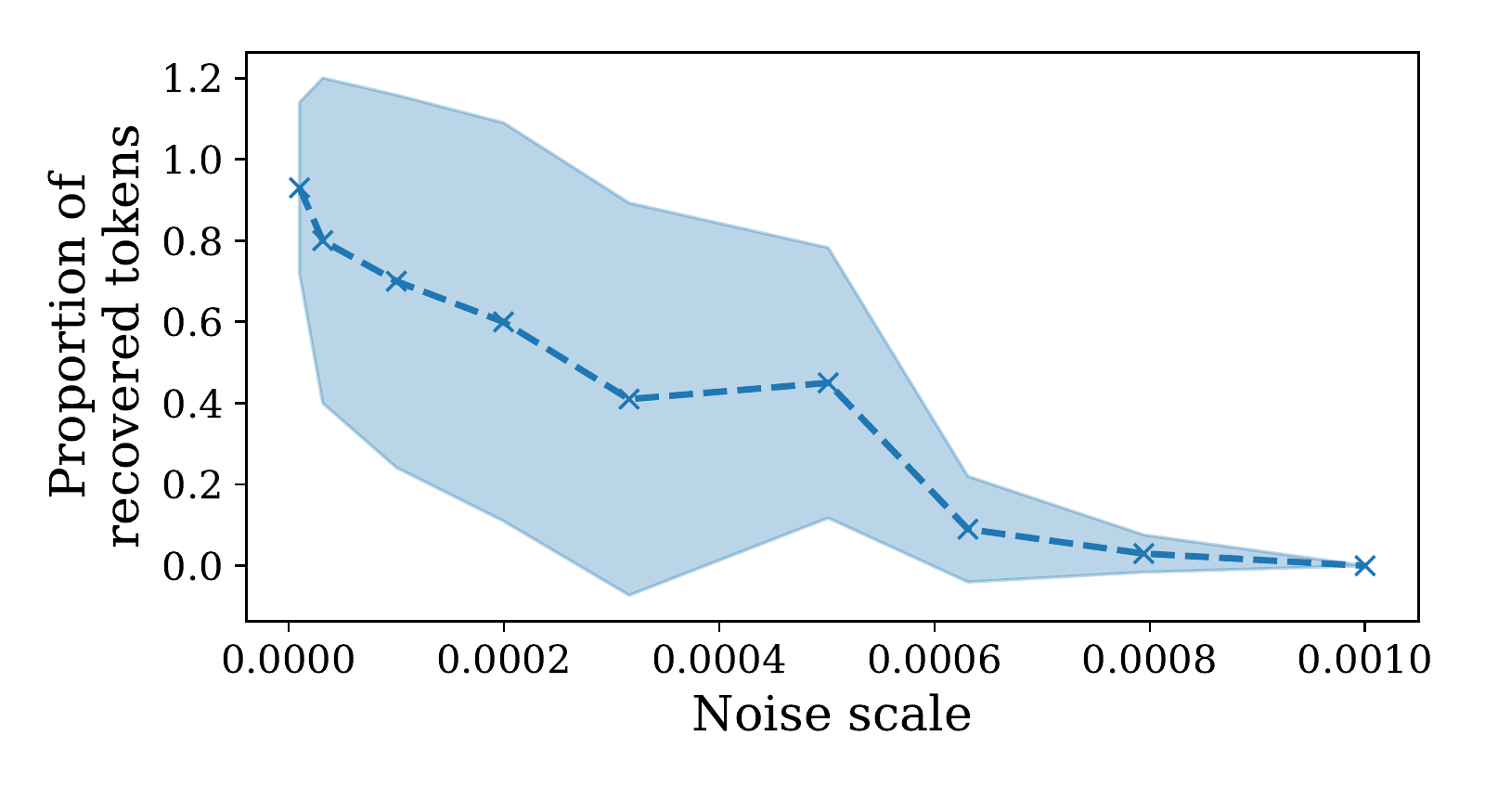}
    \caption{\textbf{Textual Data Extraction under Noise.} Extraction performance under noise for DDP from language model on the IMDB dataset. Extraction remains successful, even in presence of noise. Occasional drops in performance occur because of near-zero gradients resulted from correct data classification, \ie data points with very low original loss. Error bars correspond to a single standard deviation.}
    \label{fig:app-dp-language}
\end{figure}

\myparagraph{Textual Data.}
\label{sub:textual_data_extraction_under_dp}
For the text classifier on IMDB, we initialize the weights of the embedding layer with a random uniform distribution (minimum=0.0,maximum=1.0) to create the inputs for the fully-connected layer, following~\cite{boenisch2021curious}.
We then adversarially initialize this fully-connected-layer’s weights with the trap weights to perform extraction of the embeddings and invert the embeddings back to tokens using a lookup dictionary. 
In the presence of noise introduced for DDP, the extracted embeddings are slightly noisy.
To overcome this, in presence of noise we perform the lookup by searching for the token with the closest embedding measured through the $\Ltwo$ distance.  
\Cref{fig:app-dp-language} shows performance of a single mini-batch language extraction in presence of DP. Just as with image data, here an attacker is capable of extracting the original sentence of the \users, despite the applied noise. 
We do observe however that there is stochasticity involved---when parametrization does well on the data point by default, extraction gets low performance since the received gradient has an extremely low magnitude and the corresponding signal gets dominated by the noise. We turn to this phenomenon in the next section. 
Supplementary results for the two additional text datasets can be found in Appendix~\ref{app:exp_text}.
\\

\noindent To summarize the results on image and textual data, we find that:
\begin{itemize}
    \item The \namenoformat~\cite{boenisch2021curious} cause input data-diversity and redundancy in resulting gradients, which can be used to cancel out some of the applied noise.
    \item NLP is not safe from attacks described in this paper, despite a more sophisticated input-embeddings mapping.
    \item Despite using DDP, an attacker often can reconstruct semantic information on the individual \user data points. This is because in the presence of untrusted other \users, DDP reduces to LDP with weak privacy guarantees from the perspective of an individual \user. 
    \item As shown in \Cref{fig:DP-utility}, having \users add more noise locally, without additional improvements of the protocol~\cite{sun2020ldp} comes with a significant decrease in utility which makes the solution less practical.%
\end{itemize}

\section{Disparate Leakage over Model Gradients}
\label{sub:different_gradient_magnitudes}

Throughout our experiments, we observe that with the exact same scale of noise added to all gradients, some extracted data points have a significantly higher SNR than others.
This effect translates into different levels of semantic similarity in the extracted data with respect to the original data as we show in \Cref{fig:dp-before-cluster}.
In this section, we explain this observation and sketch how it can be leveraged by the adversary to better extract data in the presence of~noise.

\begin{figure}[t]
    \centering
    \includegraphics[width=\linewidth]{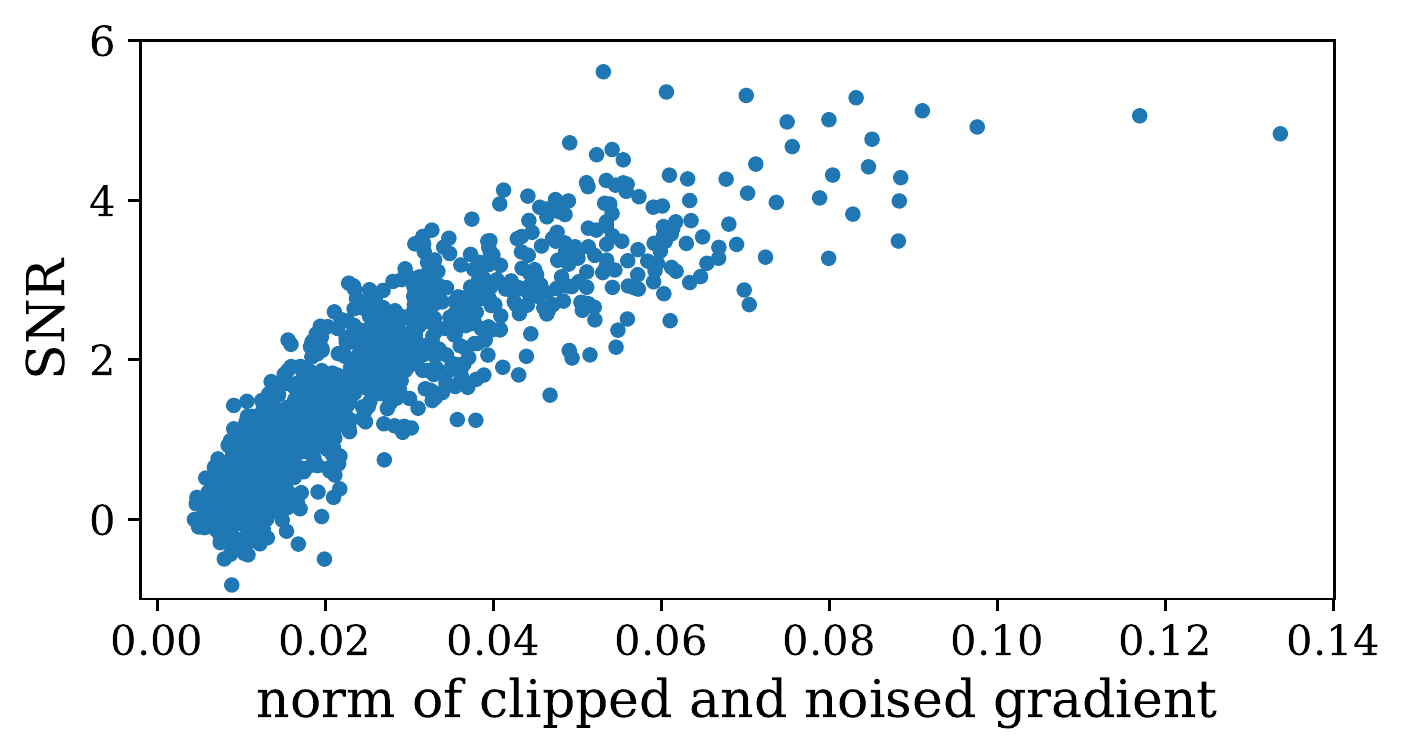}
    \caption{\textbf{Gradient Norm vs. SNR.} Norm of the clipped and noised gradients of 1000 weight rows against SNR in corresponding extracted data point, \ie the rescaled gradients. With higher gradient norms, the SNR in the extracted data increases. DDP setup: $c=1$, $\sigma=0.1$, and $\selectusers=100$.}
    \label{fig:SNR_clip}
\end{figure}

\subsection{Impact of Gradient Norm}
We find that the SNR of an extracted data point is tightly bound to the magnitude, \ie the norm, of the respective gradients.
In \Cref{fig:SNR_clip}, we depict the SNR in the rescaled clipped and noised gradients, \ie, the extracted data points, against their respective gradient norms.
The figure shows that with higher magnitude gradients, the same amount of noise has less impact on the signal, whereas, with smaller magnitude gradients, the same amount of noise largely dominates the signal.
Therefore, increased magnitude of model gradients results in an increased data leakage.

The norm of a weight row's gradients in the model depends on the model's loss.
In general, higher loss results in higher magnitude gradients, in particular for the weight rows that most contribute to the loss.
Intuitively, to increase data leakage from noisy gradients, the \cp could, therefore, manipulate the shared model to produce higher loss.
In the best case, the loss would be caused by all weight rows in the fully-connected layer used for extraction with the trap weights.
This ensures high-magnitude gradients at all the weight rows' gradients and, thereby, enables enhanced extraction at all of them.

\subsection{Global vs. Local Effect of Clipping}

\begin{figure}[tb]
    \centering
    \includegraphics[width=\linewidth, trim={4cm 7.5cm 4cm 0cm},clip]{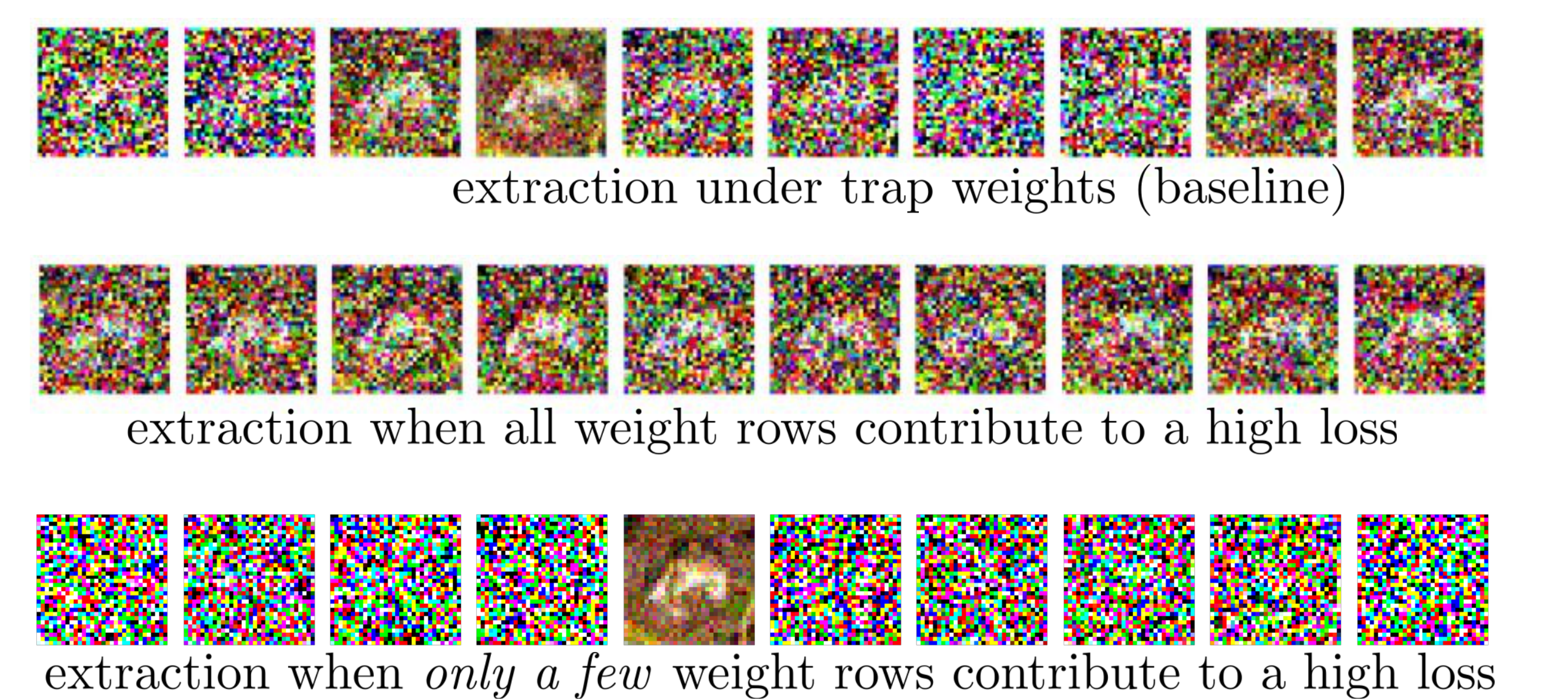}
    \caption{\textbf{Extraction Success with Additional Model Manipulations.} 
    \textbf{Row 1 (top):} Under \namenoformat only (baseline), gradients at different weight rows have varying SNRs under the same amount of added noise, depending on their magnitudes. %
    \textbf{Row 2 (middle):} When the shared model is further manipulated (\Cref{sec:reconst-with-ddp}) and all weight rows contribute equally to a high loss, their gradients will be clipped, which results in equal information loss for all of them. 
    \textbf{Row 3 (bottom):} When only a few weight rows contribute to a high loss, their gradients preserve a high magnitude over clipping, which allows for higher fidelity extraction. DDP setup: $c=1$, $\sigma=0.1$, and $\selectusers=10$.}
    \label{fig:extraction-improvement}
\end{figure}

However, in DDP, before noise addition, \users perform a clipping step, bounding the maximum per-layer gradient norm, and hence the extractable signal from a gradient update.
More precisely, clipping bounds the total norm of a model layer's gradients to the clipping parameter $c$.
If all weight rows have high gradients, their joint norm will exceed $c$, and therefore, all of them will have to be scaled down to reduce the total norm to $c$.
The effect is visualized in the middle row of~\Cref{fig:extraction-improvement}.
It shows that in this scenario, the extracted data over all gradients has a relatively low signal, which yields low-fidelity reconstruction.

Even though with DP and clipping, it is not possible to have high magnitude gradients over \textit{all} weight rows, we note that the clipping is performed \textit{globally} per-model layer.
Hence, if only a few weight rows \textit{locally} have a high magnitude gradient but all other weight rows have a low magnitude gradient, then their joint norm can be below $c$.
As a consequence, no clipping will be performed.
The effect of this scenario is visualized in the bottom row of~\Cref{fig:extraction-improvement}.
It highlights that while most gradients yield pure-noise reconstruction, a few gradients contain a very high-fidelity reconstruction of the input data.
These gradients correspond to individual neurons whose gradients were less affected by the clipping operation due to the local vs. global effect we described.
This higher vulnerability of certain neurons is desirable for improved data reconstruction under DDP.

\begin{figure}[t]
    \centering
\includegraphics[width=\linewidth]{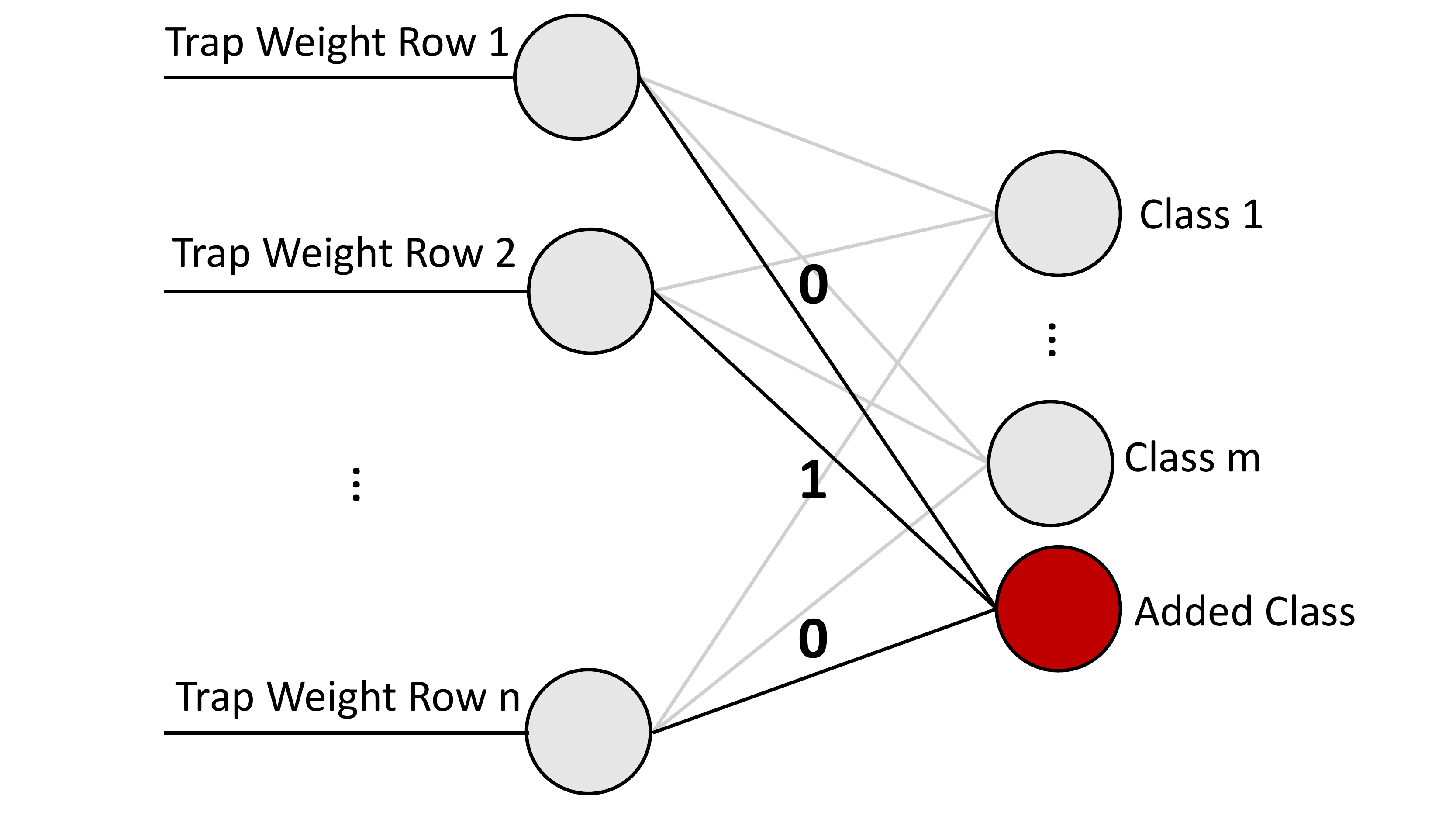}
    \caption{\textbf{Amplification of Gradient Magnitudes}. The misclassification of an example to the added class increases the magnitude of its gradient for weight rows that contribute to the loss.}
    \label{fig:app-new-manipulation}
\end{figure}
\vspace{0.7cm}

\subsection{Exploiting Global Clipping for Increased Extractability}
\label{sec:reconst-with-ddp}

The previous section highlights that the \textit{global} per-layer clipping in DP still allows \textit{local} parts of the gradients to be large.
This can be exploited for higher-fidelity data extraction from neurons corresponding to these gradients.
In this section, we sketch the idea for a possible model manipulation that gives an attacker the control on local gradient magnitudes to amplify this effect.

We base our manipulation on a fully-connected neural network with two layers.
The first layer is initialized with the trap weights for extraction and has a ReLU activation function.
The second classification layer is modified to yield high loss (without knowledge of the \user data).
Therefore, we add an additional neuron to the classification layer, \ie, an additional class that does not occur in the data distribution.
Then, we set most of the weights connecting the output of the previous layers' neurons to this additional class to very small values, \eg zero, and the weights for a few neurons' output to high values,~\eg~one.

Due to the ReLU activation of the first layer, this layer's outputs are positive.
High weights, connecting neurons to the added class in the second layer, "attribute" the loss of the misclassification to a few trap weight rows.
As a consequence, only the gradients of these few trap weight rows obtain a high magnitude (as illustrated in~\Cref{fig:app-new-manipulation} for the \textit{Trap Weight Row 2}).
All other trap weight rows (row 1 and n in~\Cref{fig:app-new-manipulation}) have low-magnitude gradients.
The overall norm of the layer's gradients will stay below the clipping parameter~$c$, hence no clipping will be applied, enabling high-fidelity data extraction from the \textit{Trap Weight Row 2}.

We instantiated the above construction with a fully-connected neural network consisting of 1000 neurons in the first layer and eleven neurons in the second layer for experimental evaluation.
The first layer's weights were initialized  with trap weights, while the second layer was initialized with a Glorot uniform distribution.
We then manipulated the weights connecting to the eleventh (added) class and set varying fractions (10\%, 30\%, 50\%, and 100\%) of them to one and the rest to zero.
Using the CIFAR10 dataset, we performed data reconstruction.

In \Cref{fig:extraction-improvement}, we visualize the extracted data. 
For the top row, all weights of the second layer are initialized at random.
In the middle row, 100\% of the weights connecting to the added class are set to one, and in the bottom row, 10\% of these weights are set to one and the rest to zero.

\begin{figure}[tb]
\centering
\begin{subfigure}[b]{0.23\textwidth}
\centering
\includegraphics[width=\linewidth]{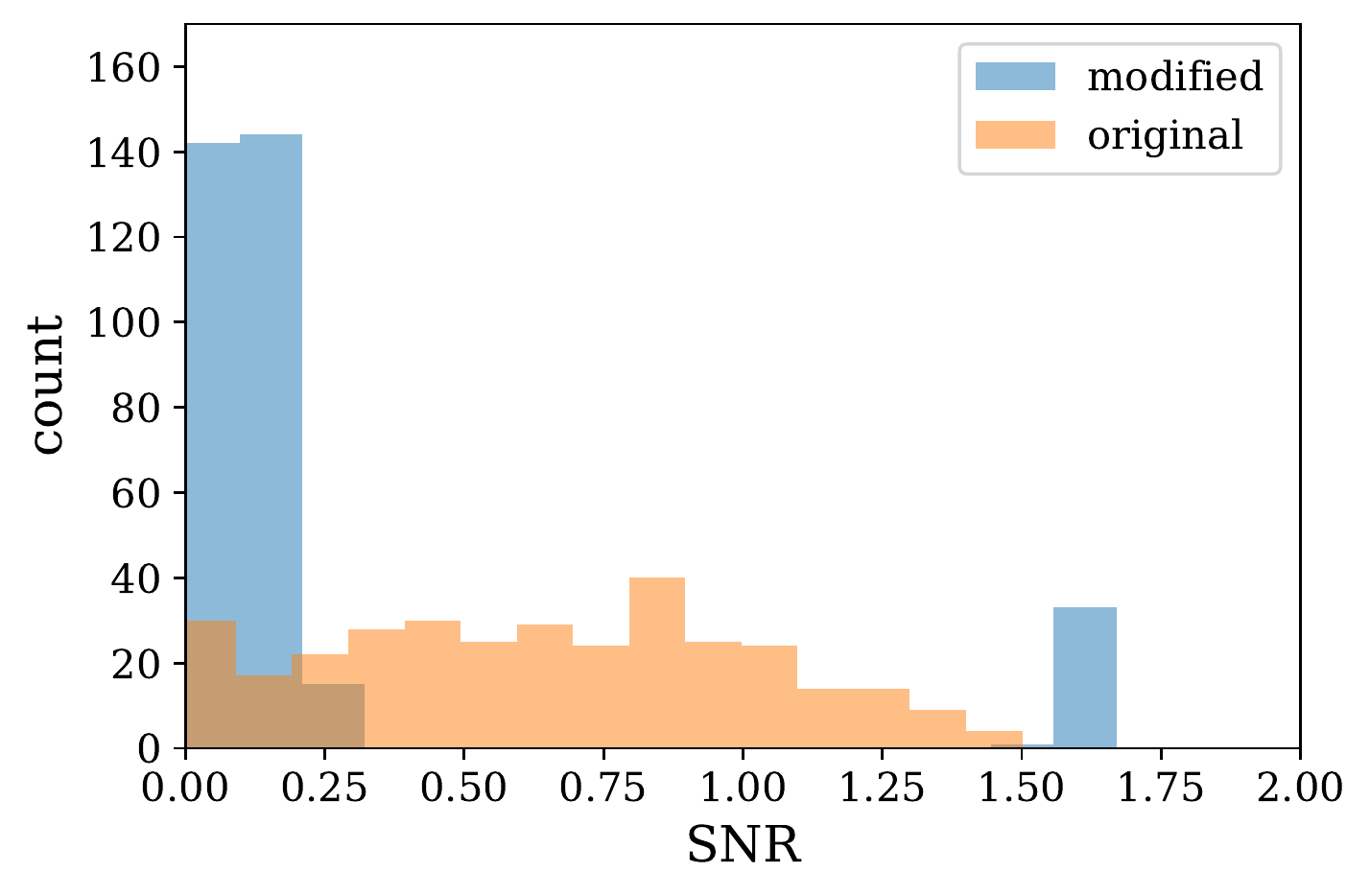}
\caption{100}
\end{subfigure}     
\begin{subfigure}[b]{0.23\textwidth}
\centering
\includegraphics[width=\linewidth]{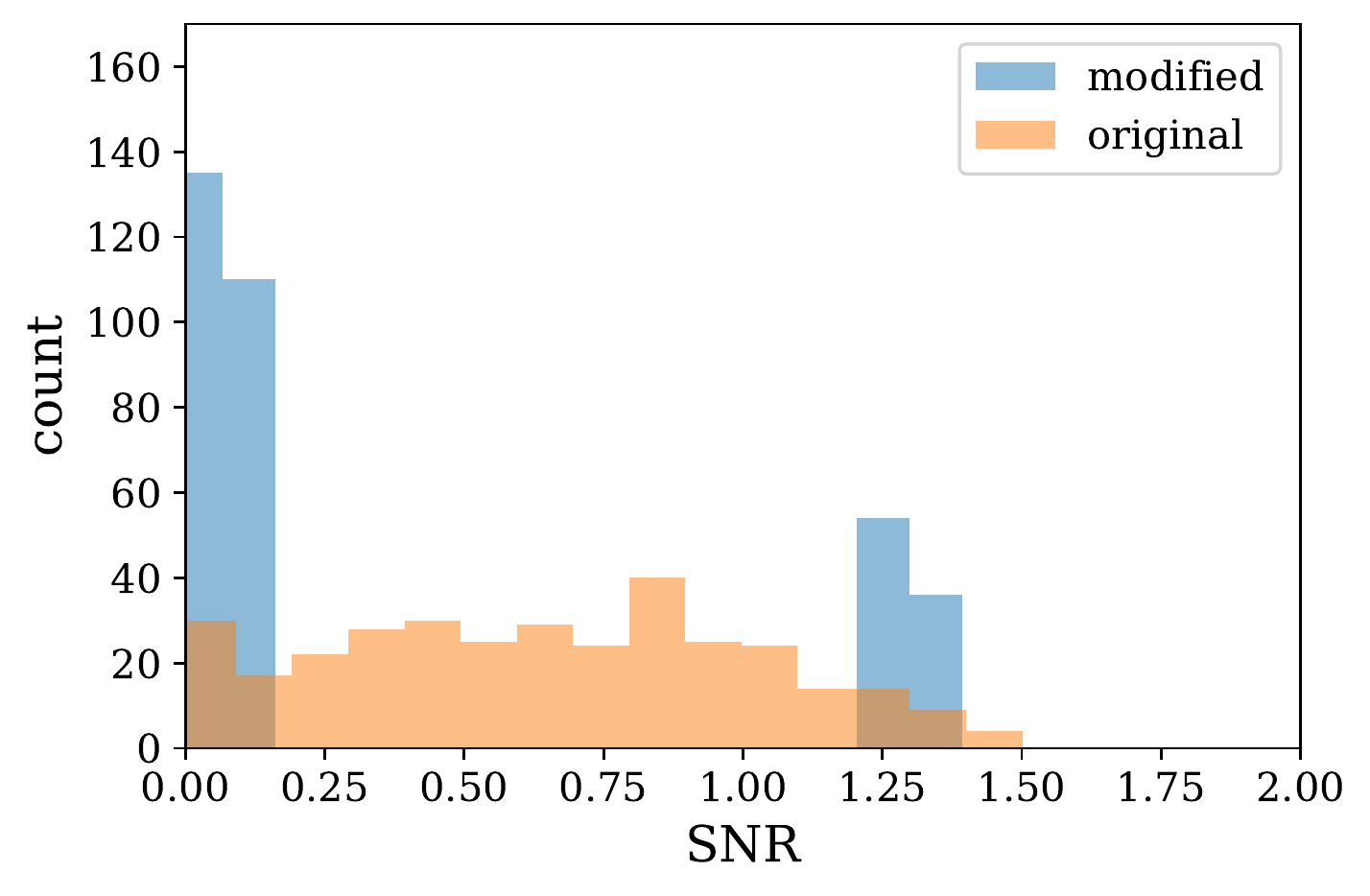}
\caption{300}
\end{subfigure}  
\hfill
\begin{subfigure}[b]{0.23\textwidth}
\centering
\includegraphics[width=\linewidth]{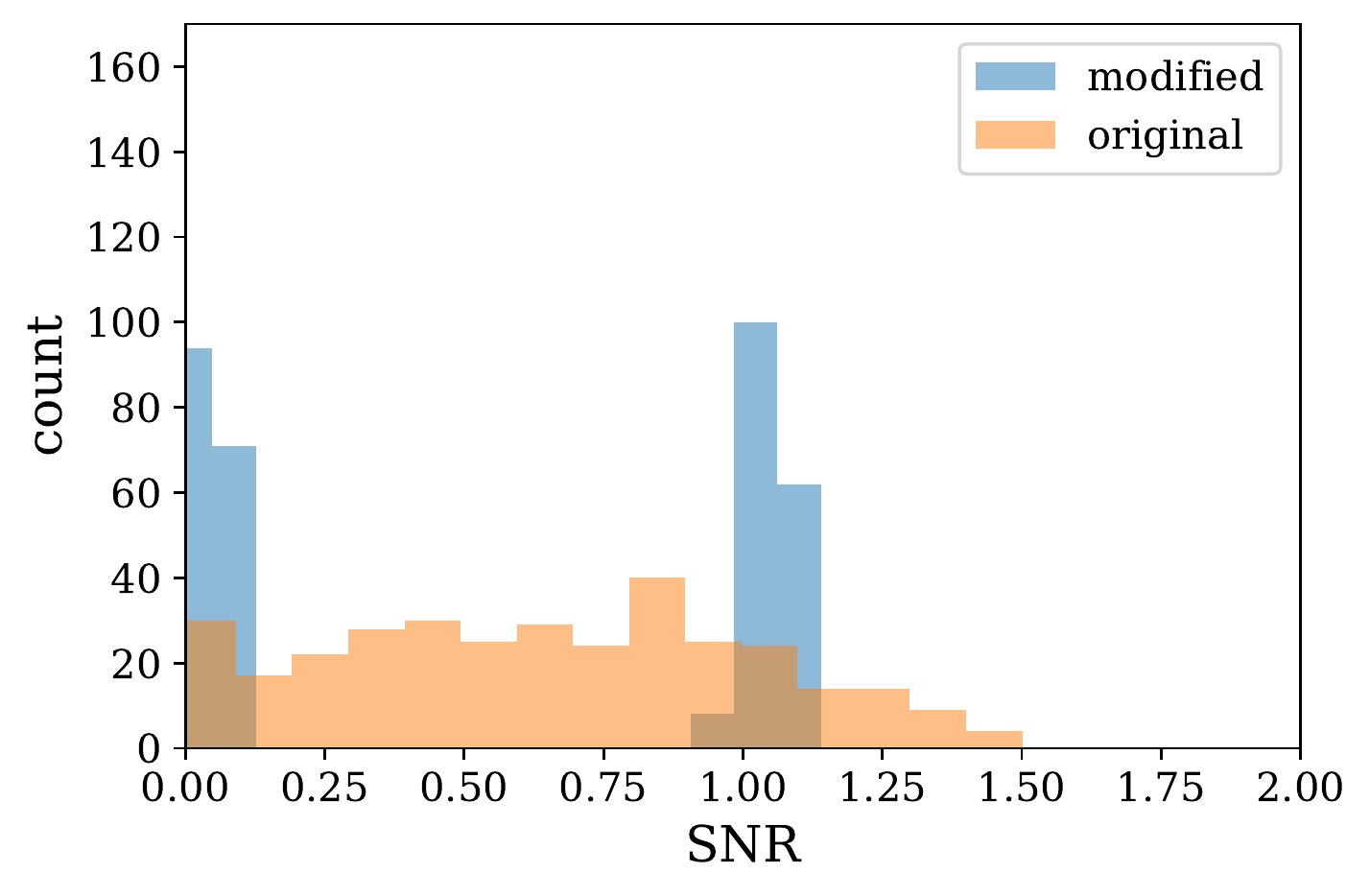}
\caption{500}
\end{subfigure}     
\begin{subfigure}[b]{0.23\textwidth}
\centering
\includegraphics[width=\linewidth]{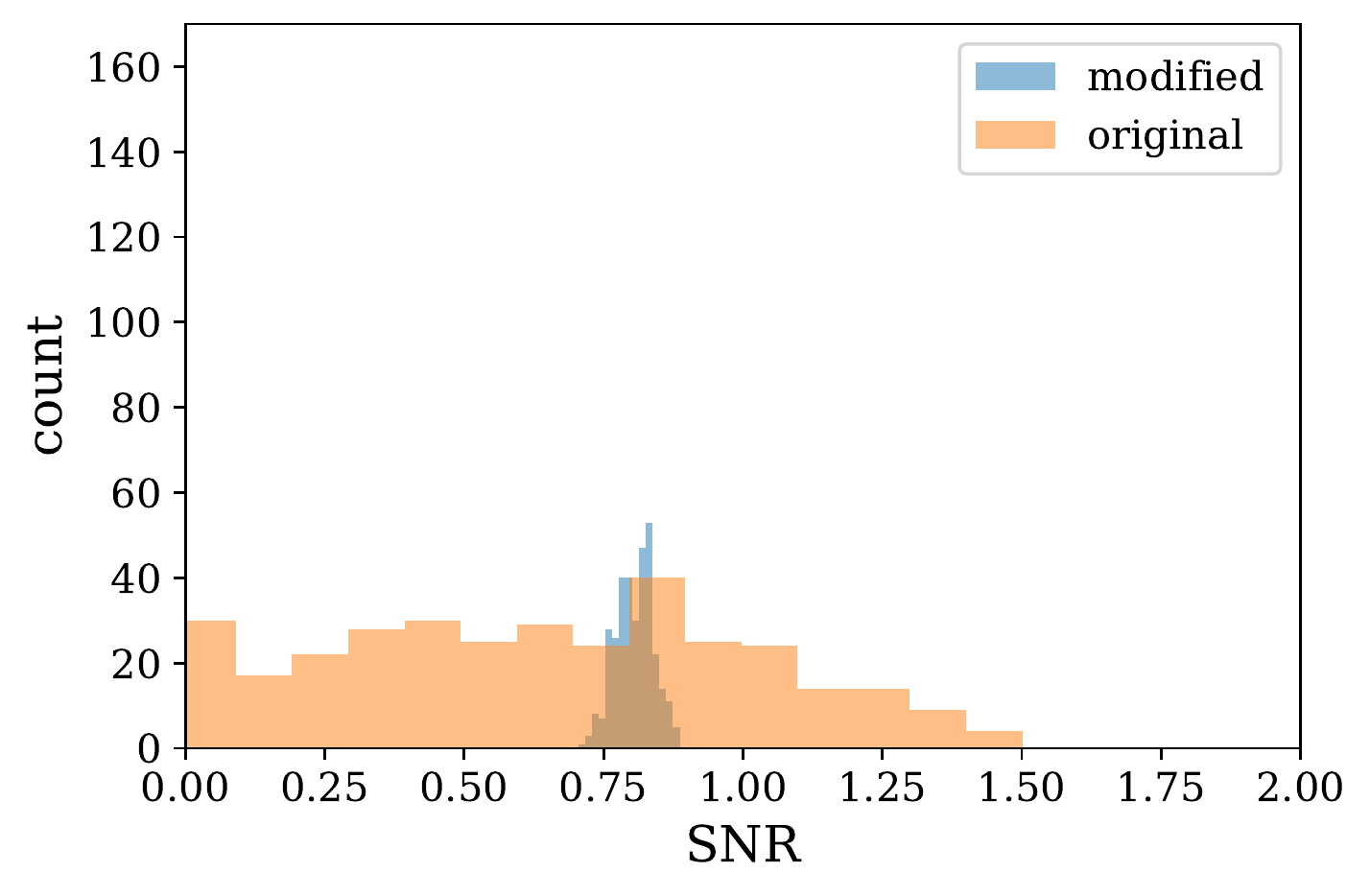}
\caption{1000}
\end{subfigure}  
\caption{\textbf{SNRs of Rescaled Clipped and Noised Gradients,} \ie the extracted data points. The first fully-connected layer used for extraction consists of 1000 neurons. The respective weight rows are initialized with our \namenoformat. The second layer is manipulated by adding an additional class neuron, and setting \{100, 300, 500, 1000\} of the 1000 weights that are connected to this neuron to one. The remaining weights that go to this neuron are set to zero. The \textit{original} baseline consists in randomly initialized weights for second fully-connected model layer. Noise with scale~$0.001$ is added to all gradients; clipping parameter~$c=1$.
When fewer weight rows (in this case 100) contribute to the high loss, the data points extracted from the respective rescaled gradients have the highest SNR ($>1.5$), which allows for higher fidelity extraction (compare to Row 3 (bottom) in~\Cref{fig:extraction-improvement}).
}
\label{fig:influence_of_ones}
\end{figure}

We furthermore measured the SNRs of the extracted data and depict the results in \Cref{fig:influence_of_ones}.
The results are consistent to the observations from \Cref{fig:extraction-improvement}:
When few weight rows contribute to the high loss, their respective rescaled gradients, \ie the extracted data points, have a higher SNR.
This allows for higher-fidelity extraction.
In contrast, when all weight rows contribute equally to the high loss, all their rescaled gradients have a similar (lower) SNR.
This is because of all their gradients being (equally) affected by the clipping.
In general, the more weight rows contribute to a high loss, the lower their individual SNRs.

Our two-layer construction naturally integrates with other architectures starting with convolutional and embedding layers described in this work.
We leave fine-tuning and the extension of our construction to architectures that end with more than two fully-connected layers for future work.
However, the approach highlights that even when DDP is in place, the \cp can initialize a model to increase the likelihood of reconstructing points with high fidelity. %

\section{Related Work}
\label{sec:RelatedWork}

We survey related work on privacy attacks against model gradients, in particular in the setup of FL.

\myparagraph{Passive Attacks against Vanilla FL.}
\label{sub:passive_extraction}
Phong~\etal~\cite{Phong.2017Privacy} were the first to show how gradients leak information that can be used to recover training data at single neurons or linear layers. Recent work~\cite{Phong.2017Privacy,Wang.2019Beyond,Zhu.2020Deep,Zhao.2020iDLG,Geiping.2020Inverting,Yin.2021See} proposed exploiting this leakage for data reconstruction, for example, through Generative Adversarial Networks \cite{goodfellow2014generative} (GANs), or by solving a second order optimization problem. %

\myparagraph{Active Attacks against Vanilla FL.} 
Melis~\etal~\cite{Melis.2019Exploiting} proposed membership~\cite{Shokri.2017Membership} and property inference~\cite{ateniese2015hacking} attacks based on periodically analyzing the model updates in an FL setup.
They consider both passive and active \attackers in vanilla FL.
Nasr~\etal~\cite{nasr2019comprehensive} assume active \attackers (both \users and \cp) for membership inference.
Similarly to Melis~\etal, their \attackers do not directly alter the shared model, but rather manipulate it through model updates (\users through  gradients, and the \cp by modifying the aggregated model update).
Recent work~\cite{fowl2021robbing,boenisch2021curious,wen2022fishing}, have shown that manipulating the shared model allows a malicious \cp to extract \user data perfectly from the model gradients.
In contrast to all these attacks that consider vanilla FL, our work attacks FL with additional extension for dedicated privacy protection through DDP and SA.
As we discussed in \Cref{sub:ddp_reconstruction}, the extraction attacks for vanilla FL can be integrated into our attack flow for improved data extraction.
We show this using \cite{boenisch2021curious}'s trap weights.

\myparagraph{Attacks against Hardened FL.} 
To our knowledge, the only prior work in this vein is due to Pasquini~\etal~\cite{pasquini2021eluding}. 
This work noticed that in \SA the \cp has the ability to dispatch inconsistent models to different \users, and used this to circumvent \SA entirely.
As we note in \Cref{sub:circumvent_sa}, their attack against \SA is not able to circumvent DDP since even when all \users's models but the target \user's model produce zero gradients, benign \users would still add their share of noise.
Thereby, privacy guarantees through DDP can still be achieved.
Finally, while Pasquini~\etal focuses on leveraging a specific capability of the \cp to circumvent a specific mechanism, we systematically study the \cp's capabilities arising from FL's pervasive centralization, and consequently offer a rich breadth of contributions over this work.
We suggest a simpler and more powerful attack that relies on sybils introduced by the \cp to circumvent \SA, compose this attack with other attacks relying on other capabilities to extract individual data points (the attack by \cite{pasquini2021eluding} only extracts updates, not individual \user data points) and attack a variant that also includes DDP.%

\section{Discussion: Privacy-Preserving FL}
\label{sec:recommendations}

In this section, we discuss the following three main questions:
\begin{itemize}
    \item \textit{Q1: What is the core reason behind FL's vulnerability to privacy attacks as the one of this work?}
    \item \textit{Q2: How can the vulnerability be fixed?}
    \item \textit{Q3: What privacy guarantees can, as of now, be provided to FL users?}
\end{itemize}

\subsection{Q1: What is the Root Cause of Vulnerability?}

We posit that \emph{the root cause of FL's vulnerability to attacks, like the one introduced in this work, is the \textbf{power imbalance} in the centralized design of FL}.
At its core, FL is a highly centralized protocol where the \cp makes the final decisions. 
For example, DDP+\SA is commonly regarded as a privacy-enhancing solution in modern FL, adding local noise and decentralizing the aggregation step of the original design, however, the \cp can circumvent the defense by controlling \users and manipulating the shared model.
In this paper, we demonstrate that even though the protocols behind DDP+\SA and their fundamental cryptographic primitives are correct, the underlying assumptions are not met when the \cp is malicious.

We detail the factors enabling the different aspects of the attack presented in this work:
\begin{enumerate}
    \item \label{itm1} The \cp is able to control a fraction of \users.
    \item \label{itm2} The \cp provisions \users for participation in each protocol-round.
    \item \label{itm4} The \cp holds the power to manipulate the shared model.
    \item \label{itm5} The \users have no inherent way of verifying each other's correctness.
    \item \label{itm6} The \users cannot meaningfully validate model updates.
\end{enumerate}

Decentralization can indeed help balance the power disparity, but it should be introduced very carefully and consider the system as a whole~\cite{pasquini2022privacy}. Next, we elaborate more on decentralization and other methods that can address the vulnerability.

\subsection{Q2: How to Fix the Vulnerability?}

We analyze how to fix the vulnerability by considering the following three approaches: (1) decentralization, (2) verification on the \user-side to decrease trust assumptions made about the \cp, and (3) application of external hardware and (cryptographic) protocols that implement guarantees under the presence of an untrusted \cp.

\myparagraph{Decentralizing FL.}
Several approaches were introduced to decentralize parts of FL, or the entire protocol.
We survey them and discuss their practical applicability.

Decentralized methods for selecting a protocol round's participants are available~\cite{girgis2021differentially, balle2020privacy, yang2021anarchic, shayan2021biscotti}. 
For example, mechanisms where the \users perform a self-sampling, such as~\cite{girgis2021differentially}, have been put forward.
These allow \users to decide about their participation and, thereby, reduce the power of the \cp.
Anarchic FL~\cite{yang2021anarchic} goes even further and lets \users choose an arbitrary point in time to update the shared model with fully-individualized training parameterization; the \cp observes \user updates directly, but each user can individually determine their required privacy level and act accordingly (for example, train on a lot of data, or add a lot of noise).
This does not only mitigate the attack vector where the \cp samples a target \user along with sybil devices.
If the \users implement enough protection locally, the \cp cannot extract their private data.
Yet, such mechanisms come with a significant increase in operational complexity and are faced with the major challenge of motivating users to provide truthful private data, compromising overall system~utility.

An interesting decentralized-FL attempt, Biscotti~\cite{shayan2021biscotti}, addresses the issue of sybil attacks by offering a block-chain based protocol that selects users in a decentralized fashion based on past behavior that appears to demonstrate honesty. This is called Proof-of-Federation (PoF), and honesty is indicated by contributing updates to the model that appear close to many other updates (and are thus assumed to be of high quality).
However, users can try to appear honest and yet act maliciously in ways that will not affect the update-quality metric, for example, by not performing their role when they are supposed to noise or verify updates (roles for round participants in the Biscotti protocol which have no performance quality metric). Biscotti would have to be extensively audited for vulnerabilities to this and other attacks before it can be safely and widely deployed. Ultimately, the designs of these decentralized FL systems are very different from FL's original design, and usually from each other's designs. At the time of writing, we are not aware of any prominent real-world FL deployments that adopt such decentralization.

\myparagraph{\User-Side Verification.}
Alternatively to decentralization, or in addition to it, we can try to reduce the trust that \users have to place in the \cp by performing verification on the \user side.

First, we discuss the verification of the shared model.
Giving users the ability to verify the shared model's integrity can prevent malicious manipulations against it, such as the ones of the trap weights that we integrated into our attack flow.
Unfortunately, without any changes to FL, there is no full-proof way to distinguish weight manipulation from model weights resulting from legitimate previous training. 
\Users have no insights on the data of other \users, updates are affected by stochasticity including sampling~\cite{Shumailov.2021Manipulating}, non-deterministic hardware elements~\cite{jia2021proofoflearning}, and large-scale FL applications with control flow mechanisms~\cite{bonawitz2019towards} that do not sample every user in every round~\cite{ramaswamy2020training}.
Ultimately, this makes it impossible to track how the model evolves over time since a \user gets only intermittent views of the shared model. 
Given these difficulties, we might consider solutions that modify FL to make manipulation-detection easier.

Second, we discuss the option of \users verifying other \users.
Privacy guarantees in DDP result from all \users adding their share of noise to protect the aggregate of all gradients.
If only one \user does not add their share of noise, the claimed theoretical overall privacy guarantees cannot be reached.
Our attack exploits this effect and exposes the gradients of a target \user by omitting noise addition of all other \users sampled in the protocol round.

To prevent this attack vector, \users have to verify that other \users calculate their gradients correctly and add the correct amount of noise.
However, due to the centralization in standard FL protocols, \users do not have direct communication channels with each other. 
The central party acts as an intermediary in their interactions.
As an alternative, FL could be deployed within a Public Key Infrastructure (PKI)~\cite{bell2020secure}.
However, a \cp that behaves semi-honestly is required during the key collection phase of PKI.
Hence, \users rely again on their trust in the \cp while nothing in the protocol prevents this \cp from acting maliciously and registering their introduced sybil devices to the PKI prior to training.

In fact, protocols like SA rely on the assumption that \users participating in the protocol are actual \users and no sybil devices controlled by the server~\cite{bell2020secure}.
Yet, nothing in the integration of the protocol into FL verifies or enforces this assumption.
As a consequence, \users who do not trust the \cp will have to verify that other \users participating with them in a protocol round follow the protocol, correctly calculate their gradients, and add their share of noise.
Verifying correct gradient calculations without \users having to reveal their private data to each other, can, in principle, be implemented through zero knowledge proofs (ZKPs) where \users commit to their private data and prove correct gradient calculation.
However, in our attack, the sybil devices are controlled by the \cp. 
As a consequence, even when they compute correct gradients, the server will be able to subtract these from the aggregate gradient to obtain the target \user's gradients and conduct extraction as described in this work.
The same argument can be applied to correct noise addition.

This motivates the need for \users to verify that other \users are no sybil devices.
Prior sybil device detection in FL relies on analyzing gradients~\cite{fung2020limitations} under the assumption of a non-malicious \cp.
This approach fails to prevent our attack where sybils are controlled by the \cp and can, as mentioned above contribute meaningful gradients, as long as the \cp can subtract these from the aggregate.
Yet, \textit{if} \users in FL have a way to confidently determine if other \users are sybil devices, and \textit{if} \users have a chance to refrain from contributing to the protocol under their presence, our attack can be mitigated.

Finally, it is desirable to enable \users to verify the whole FL application and its local execution.
This allows them to make sure that their local client handles their data correctly, adds enough local noise, and is not manipulated by the \cp.
However, in modern FL ecosystems the FL client applications are proprietary software encapsulated in dedicated partitions~\cite{privateComputeCore} largely inaccessible to \users. 
Additionally, the developers of verification software are usually the same entity acting as the \cp, bestowing it with even more power. Therefore, we recommend that \emph{applications of purportedly privacy-preserving protocols should be open-source} so that they are available for audits and verification by the community.

\myparagraph{Hardware and Protocol Support.}
As the last approach to fixing the vulnerability of FL, we discuss the support of dedicated hardware and (cryptographic) protocols to implement guarantees for the \users.

By relying on protocols that are based on trusted execution environments (TEE), \eg~\cite{mo2021ppfl} the \cp can be prevented from manipulating the shared model.
This reduces the success of data extraction as the one underlying our attack.
Additionally, to make sure that \users always receive a well-controlled and no a manipulated model, we suggest releasing the shared model publicly, for example, in a block-chain.
This makes it impossible to manipulate and change a shared model after release.

A drawback of this solution is that it offers outside attackers access to several intermediate model states.
We argue that this is not too restricting, though, since the shared model is also sent out to a few hundreds or thousands of \users during each round. 
Hence, there exists the possibility of the internal model states being leaked, anyways.
Yet, \cite{boenisch2021curious} has shown that even non-manipulated ML models, under certain conditions, leak their private training data.
Moreover, when it comes to TEEs, they are prone to side-channel attacks, \eg~\cite{jauernig2020trusted}.
Hence, we conclude that even under such hardware protection some privacy risk for \users remains.

Homomorphic encryption (HE) is a candidate solution to protect \user gradients against leakage to a malicious \cp.
However, existing instantiations~\cite{zhang2020batchcrypt} do not prevent our attack since, for efficiency in training, \users decrypt the aggregated gradients and apply them to their (unencrypted) local model.
In our attack, the \cp controls sybil devices, hence, it obtains access to the decrypted aggregate and can extract the target \user's gradient by subtracting the sybil devices' contributions.

Finally, a cryptographic protocol that always adds enough noise to \user gradients in an aggregation step would mitigate the privacy leakage of the attack presented in this work.
The protocol should offer DP, but without making any questionable trust assumptions on other \users. 
We envision a secure multiparty computation (SMPC) protocol that performs update aggregation, much like \SA, but also ensures that the output is added with a sufficient amount of noise to implement DP before it is dispatched to the \cp. As long as within this protocol \users do not learn about each others' inputs, and the \cp only learns the aggregated (and noised) output, the protocol may be able to offer a comparatively favorable privacy-utility trade-off.
To the best of our knowledge, so far, no protocol for jointly adding sufficient amounts of noise to \user gradients in SMPC exists and given the gradients' high dimensionality, the costs of any such approach will most likely not be practical, yet.

\subsection{Q3: What Users Can Do Today?}

Let us assume that a \user wishes to attain DP guarantees, and with good reason, as DP is the most viable and protective form of privacy guarantee in use today. 
Our study of FL \rev{under \SA and DDP} demonstrates that the mere inclusion of a DP mechanism does not necessarily provide protection if this mechanism makes assumptions that are inaccurate in the context of the given system.%

We currently see two promising directions for \users to attain strong privacy guarantees.

\myparagraph{Local Differential Privacy.} The first one consists of implementing their full privacy protection locally, without trusting any other participant of the protocol, neither the \cp nor other \users to contribute to their protection.
One way to implement such protection is LDP with conservative parameterization, and while ensuring that data point reuse is accounted for and prevented when needed~\cite{thudi2022bounding}.
While LDP comes at cost of the utility of the shared models, approaches to successfully improve the trade-offs exist for FL deployments with large numbers of participants~\cite{sun2020ldp}. 
We think that improving LDP to desirable privacy-utility trade-offs is a promising  direction.

\myparagraph{FL Protocols with Trusted \Cps.}
The second option for \users to obtain privacy guarantees is to opt-out of FL altogether, if they do not trust the \cp, potentially while trying to hide this, for example, by providing randomized ``garbage'' updates to the \cp. 
However, this approach comes with a corresponding utility cost and undermines the purpose of collaborative learning to produce a performant model that fits many diverse individuals. %

Finally, these options are only available if \users can control the local FL application software, which is usually not the case (unless an opt-out option is provided).

\section{Conclusion}
\label{sec:conclusion}
Truly privacy-preserving ML must defend itself from attackers that are malicious and hence do not follow protocol.
In this work, we presented a highly efficient data reconstruction attack against FL in a strongly protected deployment, namely with \SA and DDP.
\rev{
Our attack adds sybil devices to the pool of \users who are provisioned and sampled by the \cp. 
Such sybils return model updates that are known to the \cp, thus can be easily subtracted from the total aggregate to expose model updates provided by targeted individual users.
By including trap weights into our attack, we are able to reconstruct individual \users' data points of high quality.
}
Based on the attack's success, we analyzed the question of the minimum trust model that is required to obtain meaningful privacy guarantees for the \users.
We showed that FL can provide privacy guarantees if the \users trust the \cp, or if they rely on adequate cryptographic protocols, and if relevant additional protection methods are in place.
Most of the methods for protection aim at shifting power from the \cp to the conglomerate of \users and come with significant costs or overhead.
Therefore, such systems are \rev{often} not yet practically in place.
As a consequence, we recommend that, \rev{currently}, \users only participate in FL protocols that are orchestrated by a trusted \cp.

\section*{Acknowledgments}
We would like to acknowledge our sponsors, who support our research with financial and in-kind contributions: Amazon, Apple, CIFAR
through the Canada CIFAR AI Chair, DARPA through the GARD project, Intel, Meta, NFRF through an Exploration
grant, NSERC through the COHESA Strategic Alliance, the Ontario Early Researcher Award, and the Sloan Foundation. Resources used in preparing this research were provided,
in part, by the Province of Ontario, the Government of Canada through CIFAR, and companies sponsoring the Vector
Institute.

\bibliographystyle{plain}
\bibliography{library}

\appendices
\label{app:1}

\newpage
\section{Differential Privacy and DPSGD}
\label{sec:app_DP}

DP \cite{Dwork.2006Differential} formalizes the idea that no data point should significantly influence the results of an analysis conducted on a whole dataset. 
Therefore, the analysis should yield roughly the same results whether or not any particular dataset is included in the dataset or not.
Formally, this is expressed by the following definition.
\begin{df}[$(\varepsilon, \delta)$-Differential Privacy]
\label{df:Differential Privacy}
Let $A \colon \mathcal{D}^* \rightarrow \mathcal{R}$ a randomized algorithm. $A$ satisfies $(\varepsilon, \delta)$-DP with $\varepsilon \in \mathbb{R}_+$ and $\delta \in [0, 1]$ if for all neighboring datasets $D \sim D'$, i.e. datasets that differ on only one element, and for all possible subsets $R \subseteq \mathcal{R}$
\begin{align}
    \mathbb{P}\left[A(D) \in R\right] \leq e^\varepsilon \cdot \mathbb{P}\left[M(D') \in R\right] + \delta \, .
\end{align}
\end{df}

In ML, DP formalizes the idea that any possible training data point in a training dataset cannot significantly impact the resulting ML model~\cite{Abadi.2016Deep, Papernot.2016Semi}.
One notable approach to achieve this is Differentially Private Stochastic Gradient Descent (DPSGD)~\cite{Abadi.2016Deep}.
DPSGD alters the training process to introduce DP guarantees for weight update operations, and thereby protects underlying individual data points.
Particularly, the gradient computed for each data point or a mini-batch of data points is first clipped in their $\Ltwo$-norm to bound influence.
Clipping of the gradient $g(\{x_i\}_b)$ for mini-batch $b$ of data $\{x_i\}_b$ is performed according to a clipping parameter $c$, by replacing $g(\{x_i\}_b)$ with $g(\{x_i\}_b) / \max (1, \frac{||g(\{x_i\}_b)||_2}{c})$. This ensures that if the $\Ltwo$-norm of the gradients is $\leq c$, the gradients stays unaltered, and if the norm is $> c$, the gradient get scaled down to be in norm of $c$.
After the clipping, Gaussian noise with scale $\sigma$ is applied to the gradients of each mini-batch before performing the model updates. 

\rev{
\section{Extended Experimental Evaluation}
\label{app:experiments}

We present an extended experimental evaluation on two additional image and two textual datasets for spam classification.

\subsection{Datasets}
\label{app:datasets}
We describe the additional datasets used for our extended experimentation.

\myparagraph{MNIST}~\cite{LeCun.2010MNIST} is a vision dataset consisting of 70,000 gray-scale images and  corresponding labels for ten classes.
The images are of size 28x28 pixels and depict the hand-written digits zero to nine.
The dataset is divided into a training set consisting of 60,000 images and a test set consisting of 10,000 images.

\myparagraph{ImageNet}~\cite{deng2009imagenet} is a large and complex vision dataset containing color images with 224x224x3 pixels that belong to 1,000 different classes.
The dataset contains 1,281,167 training, 50,000 validation, and 100,000 test images.

\myparagraph{Spam Mails Dataset.}
The Enron-Spam datasets~\cite{emailspam} contains 5,170 emails in English language which are tagged as legitimate or spam. 
We used the the dataset from Kaggle\footnote{https://www.kaggle.com/datasets/venky73/spam-mails-dataset}.

\myparagraph{SMS Spam Collection Dataset.}
We used the SMS Spam Collection dataset~\cite{smsspam} from Kaggle\footnote{https://www.kaggle.com/datasets/uciml/sms-spam-collection-dataset} which consists of a set of SMS messages in English language. In total, it holds 5,574 SMS messages, tagged as legitimate or spam.

\subsection{Experimental Results on Image Data}
\label{app:exp_img}

\begin{figure}[h!]
    \centering
    \includegraphics[width=\linewidth]{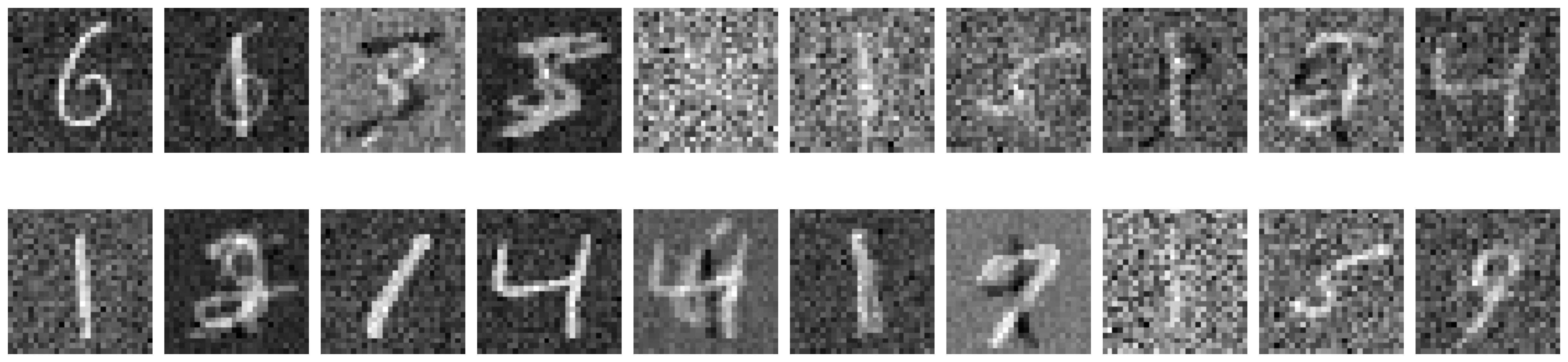}
    \caption{\rev{\textbf{MNIST: Directly Extracted Data under DDP.} Rescaled clipped and noised gradients from a mini-batch with 20 data points from CIFAR10 dataset. 
    DDP setup: $c=1$, $\sigma=0.1$, and $\selectusers=100$.}}
    \label{fig:mnist-dp-before-cluster}
\end{figure}

\begin{figure}[h!]
    \centering
    \includegraphics[width=\linewidth]{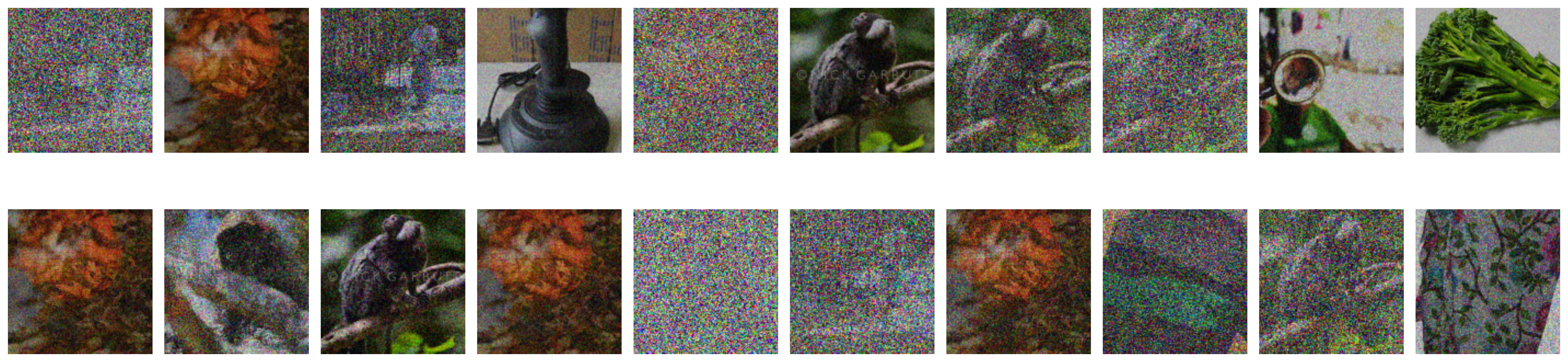}
    \caption{\rev{\textbf{ImageNet: Directly Extracted Data under DDP.} Rescaled clipped and noised gradients from a mini-batch with 20 data points from CIFAR10 dataset. 
    DDP setup: $c=1$, $\sigma=0.1$, and $\selectusers=100$.}}
    \label{fig:imagenet-dp-before-cluster}
\end{figure}

\rev{We first analyze the fraction of noisy data points that can be extracted from the gradients for MNIST and ImageNet.
With a mini-batch size of 20, for MNIST roughly ~95\% and for ImageNet roughly ~59\% of noisy data points can be extracted. 
We depict the rescaled extracted data points from gradients for MNIST in \Cref{fig:mnist-dp-before-cluster} and for ImageNet in \Cref{fig:imagenet-dp-before-cluster}.}

\rev{
\subsection{Experimental Results on Textual Data}
\label{app:exp_text}

\begin{figure}[h!]
    \centering
    \includegraphics[width=\linewidth]{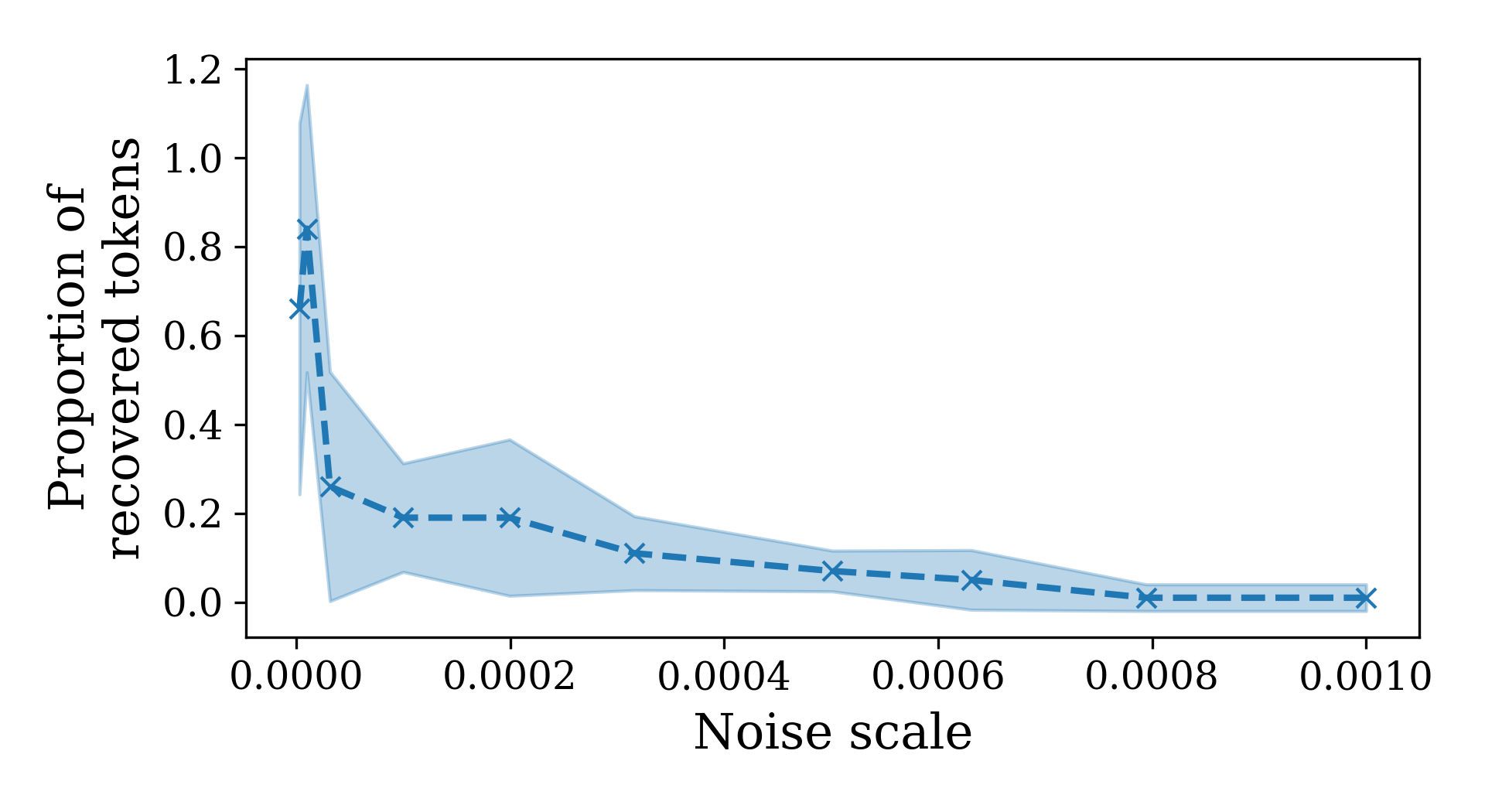}
    \caption{\rev{\textbf{Email Data Extraction under Noise.} Extraction performance under noise for DDP from language model on the Enron-Spam dataset. Extraction remains successful, even in presence of noise.}}
    \label{fig:app-dp-email}
\end{figure}

\begin{figure}[h!]
    \centering
    \includegraphics[width=\linewidth]{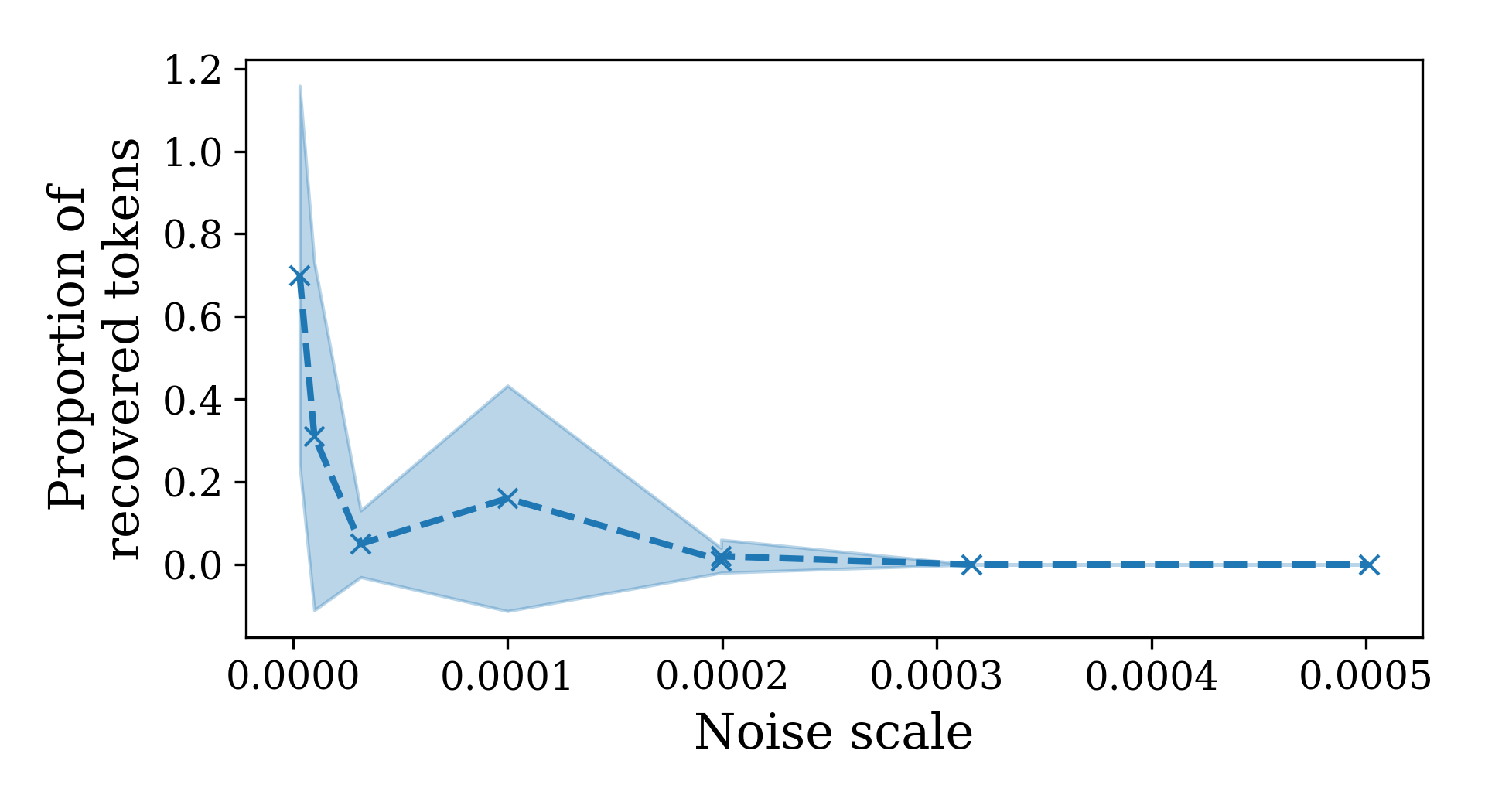}
    \caption{\rev{\textbf{SMS Data Extraction under Noise.} Extraction performance under noise for DDP from language model on the SMS Spam Collection dataset. Extraction remains successful, even in presence of noise.}}
    \label{fig:app-dp-sms}
\end{figure}

}

We present the results of extraction under noise for the Enron-Spam dataset and the SMS Spam Collection dataset in \Cref{fig:app-dp-email} and \Cref{fig:app-dp-sms}, respectively.
For both datasets we observe a degrease in the fraction of recovered tokens with increased noise scale.
The performance in the SMS Spam Collection dataset drops faster than in the email-based Enron-Spam dataset.
This effect is most likely to the limited number of token in SMS messages in comparison to emails or the reviews in the IMDB dataset.}

\end{document}